\documentclass[11pt]{article}

\newif\ifdraft
\draftfalse

\IfFileExists{lmodern.sty}{\usepackage[T1]{fontenc}\usepackage{lmodern}}{}
\usepackage[utf8]{inputenc}
\usepackage[a4paper,margin=2.5cm]{geometry}
\usepackage{amsmath,amssymb,amsthm,mathtools,bm}
\usepackage{microtype}
\usepackage{graphicx}
\usepackage{booktabs,tabularx,array}
\usepackage{enumitem}
\usepackage[table]{xcolor}
\usepackage[round,authoryear]{natbib}
\usepackage[colorlinks=true,linkcolor=blue!50!black,citecolor=blue!50!black,urlcolor=blue!50!black]{hyperref}
\usepackage[capitalise,noabbrev]{cleveref}
\ifdraft
  \usepackage[colorinlistoftodos,textsize=footnotesize]{todonotes}
\else
  \usepackage[disable]{todonotes}
\fi

\newcommand{\numerical}{\textup{[numerical]}}
\newcommand{\hypothesis}{\textup{[hypothesis]}}

\theoremstyle{definition}
\newtheorem{definition}{Definition}
\newtheorem{remark}{Remark}
\theoremstyle{plain}
\newtheorem{proposition}{Proposition}
\newtheorem{lemma}{Lemma}
\newtheorem{theorem}{Theorem}
\newtheorem{corollary}{Corollary}
\crefname{proposition}{Proposition}{Propositions}
\crefname{lemma}{Lemma}{Lemmas}
\crefname{theorem}{Theorem}{Theorems}
\crefname{corollary}{Corollary}{Corollaries}
\crefname{definition}{Definition}{Definitions}
\crefname{remark}{Remark}{Remarks}

\newcommand{\E}{\mathbb{E}}
\newcommand{\R}{\mathbb{R}}
\DeclareMathOperator{\Var}{Var}
\DeclareMathOperator{\Cov}{Cov}
\newcommand{\avg}[1]{\langle #1\rangle}
\newcommand{\dd}{\mathrm{d}}
\newcommand{\taurel}{\tau_{\mathrm{rel}}}
\newcommand{\tauint}{\tau_{\mathrm{int}}}
\newcommand{\taulocal}{\tau_{\mathrm{local}}}
\newcommand{\taudrift}{\tau_{\mathrm{drift}}}
\newcommand{\Tobs}{T_{\mathrm{obs}}}
\newcommand{\Tmin}{T_{\Theta}}
\newcommand{\Ycal}{\mathcal{Y}}
\newcommand{\KL}{D_{\mathrm{KL}}}
\newcommand{\cond}[1]{\textup{(#1)}}   
\newcommand{\T}{^{\mathsf{T}}}

\title{Top-Level Operational Equilibrium:\\ When Is the Response of a Distributed System\\ Predictable from Its Own Telemetry?}
\author{V\'ictor Casamayor Pujol$^{1,*}$\qquad Nefeli-Marina Rouska$^{1}$\qquad Boris Sedlak$^{2}$\\[6pt]
\small $^{1}$Distributed Intelligence and Systems-Engineering Lab (DISL), Departament d'Enginyeria,\\
\small Universitat Pompeu Fabra, Barcelona, Spain\\[2pt]
\small $^{2}$Distributed Systems Group (DSG), TU Wien, Vienna, Austria\\[4pt]
\small $^{*}$Corresponding author: \href{mailto:victor.casamayor@upf.edu}{\texttt{victor.casamayor@upf.edu}}\\[2pt]
\small ORCID: V.~Casamayor Pujol \href{https://orcid.org/0000-0003-2830-8368}{0000-0003-2830-8368};
N.-M.~Rouska \href{https://orcid.org/0009-0004-1272-6252}{0009-0004-1272-6252};\\
\small B.~Sedlak \href{https://orcid.org/0009-0001-2365-8265}{0009-0001-2365-8265}}
\date{\small 16 November 2026}

\begin{document}
\maketitle

\begin{abstract}
Distributed systems deployed across the edge--cloud continuum are adapted continuously: scaled, re-routed, given more or less capacity. Each adaptation rests on an expectation of how the system will respond, yet operators cannot tell from the telemetry they already collect whether that response is predictable. We define \emph{top-level operational equilibrium} (TLOE), a notion of equilibrium for a running distributed system stated relative to the controls its operator declares. TLOE has three conditions on the system's own law: stationarity with time-scale separation; sufficiency of a top-level aggregate for the controls; and time-reversibility of that aggregate. We prove that an aggregate satisfying the last two exists if and only if the minimal sufficient statistic of the controls is time-reversible, so the coarsest admissible aggregate is derived rather than chosen; that admissibility can only be lost by declaring more controls; and that the two conditions are logically independent. Under the first two conditions, the response to a small control change follows from passive telemetry through the steady-state fluctuation--response identity, with a conjugate observable fixed by the type of control: the total work in progress for a change of load, the occupancy of a stage for a change of its capacity. For chains of exponential single-server stages with Poisson input we prove that TLOE holds exactly for load control at every stable throughput, give a sufficient condition for a control on the capacity of a single stage to still admit an aggregate, and show that none exists for the first or last stage. On reference models, a per-stage choice of conjugate overestimates the early response about threefold; blocking between stages breaks both conditions weakly (1.3\% of the conjugate's variance, a 12\% bias in the static response), while a hysteretic autoscaler leaves no admissible aggregate at any of the reaction times computed. We specify a test for each condition and show that a static test of sufficiency with an estimated top-level conjugate has no power. On simulated telemetry we quantify the window lengths the prediction and the reversibility test need (a few thousand integrated autocorrelation times for the violations studied), show that the sampling interval of the telemetry is critical, and estimate the cost of a dynamic sufficiency test, which is prohibitive for weak violations but affordable for strong ones. The evidence is on reference models; real telemetry is left to future work.
\end{abstract}

\section{Introduction}\label{sec:intro}

Services deployed across the edge--cloud continuum are adapted continuously to keep their service-level objectives (SLOs) under a fluctuating environment: replicas are added or removed, traffic is re-routed, capacity is reallocated. Every such action rests on an expectation of how the system will respond. Today that expectation comes from one of two places. Either a performance model (a queueing or layered queueing model, an architecture-level model, a learned model) predicts the response, and the model must be re-estimated as conditions drift \citep{loridobotran2014review,brosig2015quantitative,guemespalau2026simulation}; or a canary experiment measures it, exposing part of the production traffic to the change \citep{basiri2019automating,warner2018canarying}. Neither tells the operator, from the telemetry the system already produces, whether the response \emph{can} be predicted at all.

The computing-continuum literature has repeatedly used the word \emph{equilibrium} for the state in which a system behaves as intended. Early work used it informally, as an operation mode tied to an infrastructure configuration and as a way to move management away from fixed thresholds \citep{casamayorpujol2021towards,dustdar2023distributed}. Later work used it for the target state of agents that act to fulfil all SLOs \citep{sedlak2024equilibrium}, gave a formal condition identifying it with SLO compliance \citep{casamayorpujol2025distributed}, and described it as predictive consistency under perturbation, measured as a Kullback--Leibler (KL) divergence between expected and observed behaviour \citep{rouska2025equilibrium}. None of these notions relates equilibrium to time-reversibility or to the response of the system, and none can be checked on a running system without either declared goals or a designed perturbation.

Statistical physics offers a precise notion, and a result that makes it operational. In any Markov steady state, the linear response of an observable to a small change of a parameter equals the time derivative of its correlation with a \emph{conjugate} observable, the derivative of the logarithm of the stationary law with respect to that parameter \citep{prost2009generalized,seifert2010fluctuation}. The identity is exact, but in general the conjugate depends on the full internal state, which no operator observes. The question this paper answers is therefore: \textbf{when can an operator predict how a system will respond to a control change from the aggregate telemetry it already collects, and how can it check, on that telemetry alone, that the prediction is valid?} In the terms above, this asks when the conjugate is a function of top-level observables.

\subsection*{Contributions}

\begin{enumerate}[leftmargin=*]
\item \textbf{A definition} (\cref{sec:definition}). TLOE is stated relative to the declared controls $\Theta$ and a coarse-graining $\Phi$. It requires stationarity with time-scale separation \cond{S}, sufficiency of the aggregate $\Phi(X)$ for the controls \cond{F}, and time-reversibility of that aggregate \cond{R}. Health is kept separate: a system is \emph{OK} when it is in TLOE and its stationary law meets the SLOs.
\item \textbf{Structure of the definition} (\cref{sec:structure}). Condition \cond{F} is closed under refinement and \cond{R} under coarsening. Hence an aggregate satisfying both exists if and only if the minimal sufficient statistic of the controls is time-reversible (\cref{thm:resolution}): the coarsest admissible aggregate is fixed by the controls. Admissibility is monotone in the declared control set: removing controls can only restore it (\cref{cor:monotone}). \cond{F} and \cond{R} are logically independent (\cref{prop:independence}).
\item \textbf{A prediction method} (\cref{sec:consequences,sec:conjugate}). Under \cond{S}+\cond{F} the response of any top-level observable to a control change follows from passive top-level telemetry. In product-form service chains the conjugate is fixed by the type of control: global (the total work in progress, WIP) for a change of load, local (minus the occupancy of stage $k$) for a change of the capacity of stage $k$. Using the occupancy of a single stage instead overestimates the early response about threefold, or predicts the wrong sign.
\item \textbf{Exact results on service chains} (\cref{sec:chains}). For a chain of $\cdot$/M/1 stages with Poisson input, total WIP is invariant in law under permutation of the stages and time-reversible (\cref{lem:tandem}, from \citet{weber1979interchangeability} and the time reversal of the chain, \citealp{kelly1979reversibility}). TLOE therefore holds exactly for load control at every stable throughput (\cref{thm:tandem}). A control on the capacity of stage $j$ admits an aggregate if $j$ is the middle stage and the multisets of upstream and downstream service rates coincide, and never if $j$ is the first or the last stage (\cref{cor:stage}).
\item \textbf{Counter-models} (\cref{sec:blocking,sec:autoscaler}). Blocking between stages breaks \cond{F} and \cond{R} at the total WIP, weakly, and leaves no admissible aggregate for the load \numerical; the violation becomes negligible as the buffer grows. A hysteretic replica autoscaler leaves no admissible aggregate at any of the reaction times computed, and the passive prediction with the total WIP is then wrong by 41\% or more of the static response (190\% when the autoscaler's reaction time is comparable to the relaxation time).
\item \textbf{Tests and their limits} (\cref{sec:tests,sec:finite}). Each condition has a test on top-level telemetry. A static test of \cond{F} with an estimated top-level conjugate has no power; only the transient response discriminates (\cref{prop:identifiability}). On simulated telemetry (1{,}000 replicates per model) we report the prediction error and the size and power of the reversibility test as functions of the window length, with confidence intervals and sensitivity checks, and estimate the cost of a dynamic test of \cond{F}, which we do not run.
\item \textbf{Bridges} (\cref{sec:kl,sec:dbn,sec:scales}). The predictive-consistency KL of \citet{rouska2025equilibrium} becomes a test of \cond{F} jointly with a hypothesised conjugate, with a quadratic small-perturbation law; the Markov blanket of the controls in a Bayesian network of the stationary law is a sufficient statistic; and sensitivities to fast controls factorise across time-scale levels \numerical.
\end{enumerate}

\paragraph{What is not new, and is cited as such.} The steady-state fluctuation--response identity \citep{prost2009generalized,seifert2010fluctuation}; product-form laws as exponential families and their score \citep{comte2025score}; the marginal-observer framework on which TLOE builds \citep{polettini2017effective,polettini2019effective}; the missing-information identity \citep{orchard1972missing}; interchangeability of $\cdot$/M/1 queues and the time reversal of open networks \citep{weber1979interchangeability,tsoucas1987interchangeability,kelly1979reversibility}, from which \cref{lem:tandem,lem:block} follow in a few lines; and the sufficiency of a Markov blanket.

\paragraph{Scope.} All results concern Markov models of the operational state, small control changes (linear response), and windows in which the system is stationary. The exact results and the simulations are on reference models: a two-stage tandem with and without blocking, multi-stage tandems, and a queue with an autoscaler. They establish what the definition does and where it fails. Whether real edge--cloud services satisfy it, and how the method performs on real telemetry, are not addressed here (\cref{sec:limitations}).

\section{Setting and background}\label{sec:setting}

The paper draws on the vocabulary of statistics, statistical physics and distributed systems. \Cref{tab:glossary} collects the terms and symbols used throughout, with the place where each is introduced.

\begin{table}[p]
\centering\small
\renewcommand{\arraystretch}{1.08}
\begin{tabularx}{\textwidth}{@{}>{\raggedright\arraybackslash}p{0.25\textwidth}X>{\raggedright\arraybackslash}p{0.12\textwidth}@{}}
\toprule
Term or symbol & Meaning & Introduced \\
\midrule
\multicolumn{3}{@{}l}{\emph{System and telemetry}}\\
Operational state $X$ & Full internal state of the system (queue lengths, jobs in flight, resource states), modelled as a continuous-time Markov chain on a countable space $S$. & \S\ref{sec:state} \\
Controls $\theta\in\Theta$ & Parameters the control plane acts on (log arrival rates, log capacities, routing weights). The operator declares $\Theta$. & \S\ref{sec:state} \\
Aggregate $\Phi$, top-level process $Y=\Phi(X)$ & The coarse quantity the operator monitors, e.g.\ the total WIP. & \S\ref{sec:state} \\
WIP $N$, occupancy $n_k$ & Work in progress: total number of jobs in the system, $N=\sum_k n_k$, where $n_k$ is the number of jobs at stage $k$. & \S\ref{sec:chains} \\
$\lambda$, $\mu_k$, $\rho_k$ & Arrival rate, service rate of stage $k$, and its load $\rho_k=\lambda_k/\mu_k$. & \S\ref{sec:productform} \\
$\pi_\theta$, $\pi^Y_\theta$, $\avg{\cdot}$ & Stationary law of $X$, stationary law of $Y$, expectation under $\pi_\theta$. & \S\ref{sec:state}, \S\ref{sec:testF} \\
$\taulocal$, $\taurel$, $\Tobs$, $\taudrift$ & Time scales of local events, of top-level relaxation, of the observation window, and of environmental drift. & \S\ref{sec:state} \\
$\tauint$, $\Delta$ & Integrated autocorrelation time of $Y$ (a window holds about $\Tobs/(2\tauint)$ independent samples); sampling interval of the telemetry. & \eqref{eq:tauint}, \S\ref{sec:finite-design} \\
Models J, B3, B2, A & Jackson tandem; tandem with a buffer of 3 or 2 at stage 2 (blocking); queue with a hysteretic replica autoscaler. & \S\ref{sec:chains}, \S\ref{sec:finite-design} \\
\midrule
\multicolumn{3}{@{}l}{\emph{Response and conjugates}}\\
$R_A(\tau)$, $C_{AB}(\tau)$ & Change of the mean of an observable $A$ at time $\tau$ after a small step in a control; stationary covariance of $A$ at time $\tau$ with $B$ at time $0$. & \Cref{prop:fdt} \\
Conjugate $B_i$ & The observable whose fluctuations predict the response to control $i$: $B_i=\partial_{\theta_i}\ln\pi_\theta$, the score of that control. & \Cref{prop:fdt} \\
Top-level conjugate $B^Y_i$ & Score of the law of $Y$; the best estimate of $B_i$ from $Y$. It equals $B_i$, up to a constant, under \cond{F}. & \S\ref{sec:testF} \\
\midrule
\multicolumn{3}{@{}l}{\emph{Definition and structure}}\\
\cond{S} & Stationarity with time-scale separation: $\taulocal\ll\taurel\ll\Tobs\ll\taudrift$. & \Cref{def:tloe} \\
\cond{F} & Top-level sufficiency: every declared control changes the stationary law only through $Y$. & \Cref{def:tloe} \\
\cond{R} & Top-level time-reversibility: $Y$ has the same statistics played forwards and backwards, so nothing circulates in a preferred direction at the top level. & \Cref{def:tloe} \\
TLOE; OK & \cond{S}, \cond{F} and \cond{R} hold; TLOE holds \emph{and} the stationary law meets the SLOs. & \Cref{def:tloe} \\
Sufficient statistic & An aggregate that carries all the information the stationary state contains about the controls; \cond{F} says $\Phi(X)$ is one. & \Cref{rem:sufficiency} \\
Minimal sufficient statistic $\Tmin$ & The coarsest sufficient statistic: it groups the states whose probability ratio does not depend on the controls. & \S\ref{sec:resolution} \\
Admissible aggregate & An aggregate satisfying \cond{F} on $\Theta$ and \cond{R} at $\theta$. One exists if and only if $\Tmin$ is time-reversible (\cref{thm:resolution}). & \Cref{def:tloe} \\
\midrule
\multicolumn{3}{@{}l}{\emph{Tests}}\\
T-S, T-F, T-R & Tests of \cond{S}, \cond{F} and \cond{R} on top-level telemetry. & \S\ref{sec:tests} \\
Size of a test & False-alarm rate: fraction of windows in which the test rejects a condition that holds (5\% by design). & \S\ref{sec:finite} \\
Power of a test & Detection rate: fraction of windows in which the test rejects a condition that is violated. & \S\ref{sec:finite} \\
\bottomrule
\end{tabularx}
\caption{Terms and notation.}\label{tab:glossary}
\end{table}

\subsection{Operational state, controls and top-level observables}\label{sec:state}

Throughout, a \emph{distributed system} is an application deployed across the edge--cloud continuum, typically a chain of services, together with the resources it runs on: the unit an operator observes through telemetry and adapts through its controls. Everything outside it, including other tenants and user demand, is its environment.

\begin{itemize}[leftmargin=*]
\item \textbf{Operational state.} $X(t)$ is the operational state of the system (queue lengths, jobs in flight, resource states) on a countable space $S$. It is modelled as a continuous-time Markov chain (CTMC) with generator $L_\theta$. Exponential timing is assumed; phase-type service keeps the state Markov at the price of a larger $S$, while general timing gives a generalised semi-Markov process (\cref{sec:limitations}).
\item \textbf{Controls.} $\theta\in\Theta\subset\R^p$, with $\Theta$ open and connected, are the parameters the operator's control plane acts on: log arrival rates, log service capacities, routing weights. The operator \emph{declares} $\Theta$; everything that follows is relative to that declaration.
\item \textbf{Top-level observables.} A coarse-graining $\Phi:S\to\Ycal$ gives the top-level process $Y(t)=\Phi(X(t))$, for example the total WIP of a service chain.
\item \textbf{Time scales.} $\taulocal$ is the time scale of service events and of local mechanisms that are faster than the top level. $\taurel$ is the relaxation time of $Y$, i.e.\ the time scale of its slowest decaying correlations. $\Tobs$ is the observation window, and $\taudrift$ the time over which the environment changes (daily load cycles, deployments). Many controllers, such as replica autoscalers, are \emph{slower} than queue relaxation; they then act as quasi-static controls rather than as fast local mechanisms (\cref{sec:scales}).
\item \textbf{A second time scale, for statistics.} The integrated autocorrelation time
\begin{equation}\label{eq:tauint}
\tauint=\int_0^\infty \frac{C_{YY}(t)}{C_{YY}(0)}\,\dd t
\end{equation}
sets the effective number of independent samples in a window, about $\Tobs/(2\tauint)$, and is the unit of window length in \cref{sec:finite}. It is not the same as $\taurel$: in the Jackson tandem of \cref{sec:chains}, $\tauint=7.6$ for the total WIP, its slowest implied time scale on a 13-bin dictionary is 9.0--9.5 at lags 0.5--16, and its autocovariance decays on a time scale of 12.7 over $t\in[100,200]$, approaching $1/(\sqrt{\mu_2}-\sqrt{\lambda})^2=14.2$, the relaxation time of the slower station \citep{morse1955stochastic}. Window requirements stated in units of $\tauint$ must be converted accordingly.
\end{itemize}

We assume $X$ is ergodic with stationary law $\pi_\theta>0$ on $S$, and that $\theta\mapsto\pi_\theta(x)$ is differentiable for every $x$. Write $\avg{\cdot}$ for expectations under $\pi_\theta$, and
\[
C_{AB}(\tau)=\avg{A(X_\tau)B(X_0)}-\avg{A}\avg{B}
\]
for the stationary two-time covariance. $L^\dagger g=\pi^{-1}L^{*}(\pi g)$ denotes the adjoint of $L$ in $L^2(\pi)$, the generator of the time-reversed stationary process.

\subsection{The steady-state fluctuation--response identity}\label{sec:fdt}

\begin{proposition}[known]\label{prop:fdt}
Let the generator be perturbed as $L_{\theta+h}=L_\theta+hL_1+o(h)$, and let $B:=\partial_h\ln\pi_{\theta+h}\big|_{h=0}$; it has mean zero, $\avg{B}=0$. Assume that the exit rates of $L_{\theta+h}$ are bounded uniformly in $x$ and in $h$ near $0$ (in a Jackson network the total exit rate of any state is at most $\lambda+\sum_k\mu_k$), so that the chain is uniformisable and its generator is a bounded operator on $L^2(\pi_\theta)$ (\cref{app:fdt}), and that $A,B\in L^2(\pi_\theta)$. Then for every such observable $A$, the response of $\avg{A}$ at time $\tau$ after a step of size $h$ applied at time $0$ to the stationary process is, to first order in $h$,
\begin{equation}\label{eq:fdt}
R_A(\tau)=C_{A,B}(0)-C_{A,B}(\tau),\qquad\text{i.e.}\qquad \frac{\dd R_A}{\dd\tau}=-\frac{\dd}{\dd\tau}C_{A,B}(\tau),
\end{equation}
and in particular the static response is $\partial_\theta\avg{A}=\Cov(A,B)$.
\end{proposition}

This is the fluctuation--dissipation theorem (FDT) of non-equilibrium steady states \citep{prost2009generalized,seifert2010fluctuation}, who express the conjugate through the stochastic entropy $-\ln\pi$; it goes back to \citet{agarwal1972fluctuation} under detailed balance. A derivation is in \cref{app:fdt}. Conversely, if the \emph{static} identity $\partial_\theta\avg{A}=\Cov(A,B')$ holds for all bounded $A$ with some $B'$, then $\Cov(A,B'-B)=0$ for all $A$, so $B'-B$ is $\pi$-almost surely constant; since the dynamic identity implies the static one ($\tau\to\infty$), the same holds if the dynamic identity holds for all $A$. Conjugates are therefore defined up to an additive constant, which affects none of the identities in this paper, since they involve only covariances; we fix it, when needed, by centring.

The identity has two consequences that shape the rest of the paper. First, \emph{the FDT never simply fails in a steady state}: what fails is that the conjugate $B$ is simple, known and observable. Physics obtains a local conjugate from detailed balance together with local detailed balance ($B$ equals $\beta$ times the perturbing observable). Second, which observable is conjugate depends on the class of perturbation \citep{owen2020universal}. In a service chain, the class is fixed by which control the operator moves.

\subsection{Product form}\label{sec:productform}

In an open Jackson network with external Poisson arrivals of rate $\lambda$, Markov routing with probabilities $p_{kj}$, and exponential single-server stations with rates $\mu_k$, the stationary law is
\begin{equation}\label{eq:jackson}
\pi(n)=\prod_k(1-\rho_k)\,\rho_k^{n_k},\qquad \rho_k=\lambda_k/\mu_k<1,
\end{equation}
where $\lambda_k$ solves the traffic equations \citep{jackson1963jobshop}; BCMP networks share the property \citep{baskett1975open}. Taking logarithms, $\ln\pi(n)=\sum_k n_k\ln\rho_k+\text{const}$, so the family $\{\pi_\theta\}$ is an exponential family whose sufficient statistic is the vector of occupancies, and the score of a parameter $\theta$ is $\sum_k(n_k-\avg{n_k})\,\partial_\theta\ln\rho_k$ \citep{comte2025score}. Product form is a property of the stationary law only: an open Jackson network with feed-forward routing is not time-reversible at the resolution of the stations, since jobs move from stage $k$ to stage $k+1$ and never back \citep{kelly1979reversibility}. Tractability comes from product form, not from reversibility.

\section{Definition}\label{sec:definition}

\begin{definition}[Top-level operational equilibrium]\label{def:tloe}
A distributed system is in \emph{top-level operational equilibrium} (TLOE) at $\theta\in\Theta$, with respect to the declared controls $\Theta$ and the coarse-graining $\Phi$, over a window $\Tobs$, if:
\begin{description}[leftmargin=1.2em,font=\normalfont\bfseries]
\item[(S) Stationarity with time-scale separation.] $X$ is ergodic with stationary law $\pi_\theta$, the law of $Y=\Phi(X)$ is stationary over $\Tobs$, and
\[
\taulocal\ll\taurel\ll\Tobs\ll\taudrift.
\]
\item[(F) Top-level sufficiency.] For every $\theta'\in\Theta$ and every control direction $i$ there is a function $B_i=B_i(\cdot\,;\theta')$ on $\Ycal$ such that
\begin{equation}\label{eq:F}
\partial_{\theta_i}\ln\pi_{\theta'}(x)=B_i(\Phi(x);\theta')-\avg{B_i(Y;\theta')}_{\theta'}\qquad\text{for all }x\in S.
\end{equation}
\item[(R) Top-level time-reversibility.] The stationary process $(Y(t))_{t\in\R}$ under $\pi_\theta$ has the same law as its time reversal $(Y(-t))_{t\in\R}$.
\end{description}
\end{definition}

We call a coarse-graining $\Phi$ that satisfies \cond{F} on $\Theta$ and \cond{R} at $\theta$ an \emph{admissible aggregate} for the declared controls.

Conditions \cond{S} and \cond{R} concern the operating point $\theta$. Condition \cond{F} is a property of the declared family $\{\pi_{\theta'}:\theta'\in\Theta\}$ and of $\Phi$; it does not depend on $\theta$ (\cref{rem:pointwise}). The function $B_i$ in \cond{F} is defined up to an additive constant (\cref{sec:fdt}). A system is \textbf{OK} at $\theta$ if it is in TLOE \emph{and} $\pi_\theta$ meets its SLOs. Equilibrium makes the probabilistic guarantees of the system computable and its response predictable; the SLO check decides whether the guarantees are met. The two are deliberately separate: a system can be in TLOE and violate its SLOs (it is predictably slow), or meet its SLOs without being in TLOE (it is currently fast, but its response to the next adaptation cannot be trusted).

\begin{remark}[\cond{F} is sufficiency]\label{rem:sufficiency}
Since $\Theta$ is connected and $\pi_\theta>0$, \cond{F} holds if and only if $\Phi(X)$ is a sufficient statistic for $\theta$ in the family $\{\pi_\theta\}_{\theta\in\Theta}$: integrating \eqref{eq:F} along a path in $\Theta$ from a reference $\theta_0$ gives $\pi_\theta(x)=\pi_{\theta_0}(x)\,f_\theta(\Phi(x))$, the Fisher--Neyman factorisation; the converse follows by differentiating the factorisation. \cond{F} says that every control tilts the stationary law only through the top-level observables.
\end{remark}

\begin{remark}[Why \cond{F} is required on all of $\Theta$]\label{rem:pointwise}
The integration in \cref{rem:sufficiency} needs \eqref{eq:F} along the whole path. A version of \cond{F} at the single point $\theta$ is strictly weaker and does not give sufficiency: for $\pi_\theta(x)\propto g(x)\,f_\theta(\Phi(x))\exp\!\big((\theta-\theta_0)^2k(x)\big)$ with $k$ not a function of $\Phi$, the score at $\theta_0$ is a function of $\Phi(x)$, yet $\Phi$ is not sufficient. If the operator only trusts the model on a connected open neighbourhood $U\ni\theta$, \cond{F} can be required on $U$ instead; every statement below then holds with $\Theta$ replaced by $U$, and the minimal statistic $T_U$ of the restricted family is a coarsening of $\Tmin$ (fewer parameter values separate fewer states). For the Jackson families of \cref{sec:chains} the two coincide, because likelihood ratios are monomials in the rates.
\end{remark}

\begin{remark}[Weak form of \cond{R}]\label{rem:weakR}
\cond{R} has a weak form that is tested in practice: $\E[f(Y_0)g(Y_\tau)]=\E[g(Y_0)f(Y_\tau)]$ for all functions $f,g$ and lags $\tau$, i.e.\ the lagged second-moment matrix $K(\tau)_{ab}=\E[\psi_a(Y_0)\psi_b(Y_\tau)]$ is symmetric for every dictionary $\{\psi_a\}$ of functions of $Y$. The strong form implies that the relative-entropy rate between $Y$-paths and their time reversals vanishes.
\end{remark}

\begin{remark}[\cond{R} is a condition on the aggregate, not on the state]\label{rem:aggregate}
A feed-forward service chain is irreversible at the resolution of its stages: its entropy production at that resolution is infinite, because the reverse of a forward move has probability zero \citep{murashita2014nonequilibrium}. TLOE allows this. It requires reversibility only of the aggregate fixed by the controls.
\end{remark}

\begin{remark}[The roles of the three conditions differ]\label{rem:roles}
\cond{S} and \cond{F} are what the prediction needs (\cref{prop:consequences}(a)). \cond{R} is not needed by the prediction: it is part of the definition because it is what distinguishes an \emph{equilibrium} from a mere steady state. It excludes stationary behaviour that keeps circulating in a preferred direction at the top level; any detected time asymmetry of $Y$ proves such a circulation (\cref{sec:consequences}). A reader who only wants the prediction can drop \cond{R} and keep \cond{S} and \cond{F}; the price is that the resulting notion no longer separates a stationary but directed top level from an equilibrium one, and that the reciprocity and spectral statements of \cref{prop:consequences}(c)--(e) are lost.
\end{remark}

\begin{remark}[\cond{S} is partly testable, partly monitored, partly assumed]\label{rem:S}
Stationarity of $Y$ within the window and the ratio $\Tobs/\taurel$ can be tested on the window; $\Tobs\ll\taudrift$ can only be monitored across windows; $\taulocal\ll\taurel$ cannot be checked on top-level telemetry sampled more slowly than $\taulocal$, and is an assumption about what the operator integrates out (\cref{sec:testS}). Steady performance may simply not exist \citep{antoch2026experimental}, so the testable parts of \cond{S} are tested, not assumed.
\end{remark}

\paragraph{What TLOE is not.} It is not microscopic reversibility (\cref{rem:aggregate}). It is not the absence of flow: throughput can be large. It is not the terminal state of one workflow instance, which is an absorbing state rather than an ensemble; TLOE concerns the stream of instances a running service executes. It is not consensus among replicas, which is a fixed point of the control plane, not of the operational process. It is not health (SLO compliance). And it does not involve a temperature: in \cond{F} only $\theta$ and $B$ appear, and any inverse temperature is absorbed into the parametrisation of the controls.

\paragraph{Physical equilibrium as a special case.} Take $\Phi=\mathrm{id}$ and let $\Theta$ be fields that couple to the energy of a system with local detailed balance. Then \cond{F} is vacuous (the full state is always sufficient), \cond{R} is detailed balance, and \cond{S}+\cond{R} at $\Phi=\mathrm{id}$ is stationary reversible dynamics; local detailed balance then supplies the Gibbs form and the physical conjugate. TLOE departs from this case in two directions: it is relative to the controls $\Theta$, and relative to the resolution $\Phi$. It keeps the structure the response identity needs and drops what a working distributed system cannot have, namely reversibility at the resolution of individual stages.

\section{Structure of the definition}\label{sec:structure}

\subsection{What each condition gives}\label{sec:consequences}

\begin{proposition}\label{prop:consequences}
Assume \cond{S} and \cond{F}, and let $B_i=B_i(Y)$ be the conjugates given by \cond{F} at $\theta$.
\begin{enumerate}[label=(\alph*),leftmargin=*]
\item \emph{Top-level response.} For every top-level observable $A=a(Y)$ and every control $i$, the response of $\avg{A}$ to a small step $\delta\theta_i$ is $R_{A,i}(\tau)\,\delta\theta_i$ with
\[
R_{A,i}(\tau)=C_{A,B_i}(0)-C_{A,B_i}(\tau),\qquad \partial_{\theta_i}\avg{A}=\Cov(A,B_i).
\]
Every term is a two-time statistic of $Y$ alone, so \textbf{the response is computable from passive top-level telemetry}.
\item \emph{No information loss in a top-level snapshot.} The Fisher information about $\theta$ in a snapshot $Y_t$ equals that in a snapshot $X_t$: $I_{ij}(\theta)=\Cov(B_i,B_j)$. By the Cram\'er--Rao bound this limits how precisely $\theta$ can be estimated from top-level snapshots, and, by the delta method, how precisely any smooth function of $\theta$, such as a control response, can be. (The statement is about snapshots; it says nothing about the information in paths of $Y$ versus paths of $X$.)
\end{enumerate}
Assume in addition \cond{R}.
\begin{enumerate}[label=(\alph*),leftmargin=*,start=3]
\item \emph{Reciprocity between controls.} $R_{B_j,i}(\tau)=R_{B_i,j}(\tau)$ for all $i,j$ and $\tau$.
\item \emph{Symmetric lagged moments.} For every finite dictionary $\{\psi_a\}$ of functions of $Y$ (constants allowed), the uncentred matrix $K(\tau)_{ab}=\E[\psi_a(Y_0)\psi_b(Y_\tau)]$ is symmetric; if the $\psi_a$ are linearly independent in $L^2(\pi)$, $K(0)$ is positive definite and the generalised eigenvalues of the pencil $(K(\tau),K(0))$, $K(\tau)v=\kappa K(0)v$, are real.
\item \emph{Relaxation without oscillation, if $Y$ is Markov.} If $Y$ is moreover a Markov process, its generator is self-adjoint and non-positive in $L^2(\pi^Y)$, so every autocovariance $C_{a,a}(\tau)$ is a mixture of decaying exponentials with non-negative weights: the top level relaxes without oscillating, and the response to a control $i$ of its own conjugate $B_i$ is monotone in time.
\end{enumerate}
\end{proposition}

\begin{proof}
(a) is \cref{prop:fdt} with $B=B_i(Y)$. (b): the score of $\theta$ in the law of $X_t$ is $B(\Phi(X_t))-\avg{B}$, which is also the score in the law of $Y_t$; both informations equal its covariance. (c): by (a), $R_{B_j,i}(\tau)=C_{B_j,B_i}(0)-C_{B_j,B_i}(\tau)$, and $C_{B_j,B_i}(\tau)=\E[B_j(Y_\tau)B_i(Y_0)]-\avg{B_j}\avg{B_i}$ is unchanged by exchanging $i$ and $j$ when $Y$ is reversible. (d): symmetry is the weak form of \cond{R} (\cref{rem:weakR}); a symmetric pencil with positive-definite $K(0)$ has real eigenvalues. (e) is the spectral theorem for the self-adjoint generator of a reversible Markov process.
\end{proof}

\begin{figure}[t]
\centering
\includegraphics[width=\textwidth]{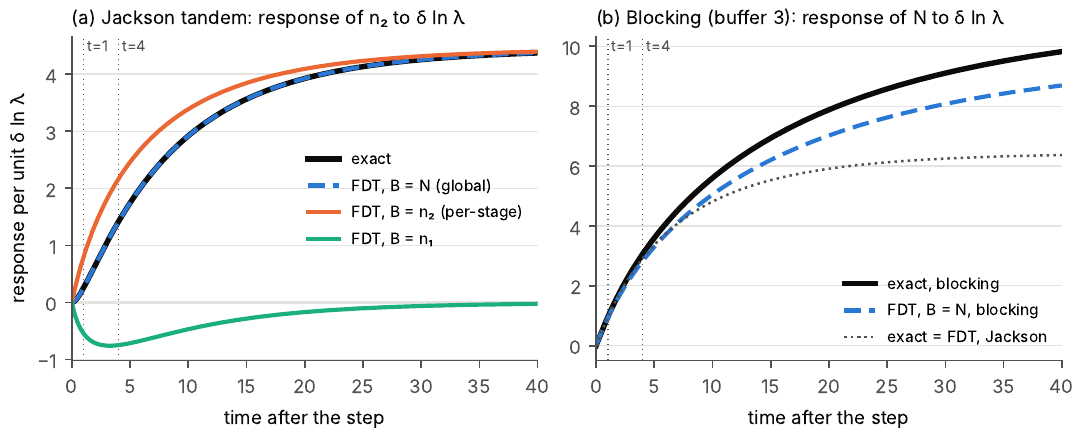}
\caption{(a) Jackson tandem: step response of $n_2$ to a change of load, exact and predicted from passive fluctuations with three choices of conjugate. The global conjugate $N$ is exact; the per-stage $n_2$ overshoots in the transient; $n_1$ has the wrong sign. (b) With a buffer of 3 at stage 2 (blocking after service), the prediction with $B=N$ falls below the exact response of $N$, increasingly with time (11.8\% in the static limit). The dotted curve is the Jackson tandem for reference, where prediction and exact response coincide. Vertical guides mark $t=1$ and $t=4$, the times of \cref{tab:conjugate}.}
\label{fig:evidence}
\end{figure}

Four cautions bound what \cond{R} gives.
\begin{itemize}[leftmargin=*]
\item \textbf{Prediction does not need \cond{R}.} \cond{S}+\cond{F} alone give the response, its timing included: $R_{A,i}(\tau)$ is the full curve. \cond{R} adds reciprocity, symmetric covariances and, for Markov $Y$, a qualitative statement about how the response unfolds (\cref{rem:roles}).
\item \textbf{\cond{R} without Markovianity does not exclude oscillation.} Every stationary scalar Gaussian process is time-reversible \citep{weiss1975time}, including one with covariance $e^{-|\tau|}\cos\tau$. In (d) the generalised eigenvalues are real but may be negative.
\item \textbf{The paper's own main example is reversible but not Markov.} The total WIP $N$ of the tandem (\cref{sec:chains}) is time-reversible (\cref{lem:tandem}) but is not a Markov process: the rate at which $N$ decreases depends on whether stage 2 is empty, so the chain is not lumpable onto $N$ \citep{kemeny1960finite}, and the Chapman--Kolmogorov identity for the two-time law of $N$ fails by 3.7\% (relative, at $t=0.5$) in model J. Part (e) therefore does not apply to it, and the monotone responses of \cref{fig:evidence} are a property of the model, not a consequence of \cond{R}. The standard Markovianity check, independence of implied time scales $t_k=-\tau/\ln\kappa_k(\tau)$ from the lag \citep{prinz2011markov}, is insensitive here: on a 13-bin indicator dictionary the slowest implied time scale of $N$ moves only from 9.0 to 9.2 over lags 0.5 to 4 (and to 9.5 at lag 16). The check is necessary, not sufficient.
\item \textbf{The implication runs one way.} An asymmetric $K(\tau)$ proves that $Y$, and hence $X$, is not time-reversible: a function of a reversible process is reversible, so coarse-graining cannot create irreversibility in the exact law. A symmetric $K(\tau)$ on a finite dictionary does not prove \cond{R}, and \cond{R} at the top level does not exclude currents in the full state, which in a service chain are always present (\cref{rem:aggregate}).
\end{itemize}

\subsection{The resolution theorem}\label{sec:resolution}

For a family $\{\pi_\theta\}_{\theta\in\Theta}$ on a countable $S$ with $\pi_\theta>0$, write $x\sim_\Theta x'$ when $\pi_\theta(x)/\pi_\theta(x')$ does not depend on $\theta\in\Theta$. The map $\Tmin$ sending $x$ to its equivalence class is a \textbf{minimal sufficient statistic} for $\theta$, unique up to a bijection of its values: it is sufficient, and it is a function of every sufficient statistic \citep{lehmann1950completeness}; \citet[Theorem~6.2.13]{casella2002statistical} give a textbook statement.

Two closure properties follow directly.
\begin{itemize}[leftmargin=*]
\item \textbf{\cond{F} is closed under refinement.} If $\Phi$ is sufficient and $\Phi=h\circ\Phi'$ for some $h$, then $\Phi'$ is sufficient.
\item \textbf{\cond{R} is closed under coarsening}, in both its strong and weak forms. If $Y=\Phi(X)$ is time-reversible, so is $g(Y)$ for every $g$, because the law of $(g(Y_t))_t$ is the image of the law of $(Y_t)_t$ under a map that commutes with time reversal.
\end{itemize}

\begin{theorem}[Resolution]\label{thm:resolution}
Let $S$ be countable, $\pi_\theta>0$, $\Theta$ open and connected, and $\theta\mapsto\pi_\theta(x)$ differentiable. Fix $\theta\in\Theta$. There exists a coarse-graining $\Phi$ satisfying \cond{F} on $\Theta$ and \cond{R} at $\theta$ if and only if the process $\Tmin(X)$ is time-reversible at $\theta$. In that case $\Tmin$ itself satisfies both, and every $\Phi$ satisfying both is a refinement of $\Tmin$.
\end{theorem}

\begin{proof}
($\Leftarrow$) Take $\Phi=\Tmin$: it is sufficient, hence satisfies \cond{F} by \cref{rem:sufficiency}, and it is reversible by assumption. ($\Rightarrow$) If $\Phi$ satisfies \cond{F} on $\Theta$, it is sufficient (\cref{rem:sufficiency}), so $\Tmin=g(\Phi)$ for some $g$ by minimality; since $\Phi(X)$ is reversible, so is $g(\Phi(X))=\Tmin(X)$.
\end{proof}

The proof is short once the closure properties are in place; we call the result a theorem because it carries the conceptual content of the definition. $\Tmin$ is defined by the whole family $\{\pi_\theta\}$, while reversibility is a property of the process at the operating point $\theta$; the theorem is a statement at each $\theta$. TLOE can therefore be restated as a single condition on the declared control family: \textbf{stationarity \cond{S}, plus time-reversibility of the minimal sufficient statistic of the controls.} The coarsest admissible aggregate is derived from the controls, not chosen; every refinement of it is also sufficient, and a refinement can be reversible only if $\Tmin$ is.

The theorem separates two kinds of failure, both met in \cref{sec:chains}. A \emph{failure at a given $\Phi$} says nothing by itself about existence: with blocking, the total WIP is no longer sufficient for the load (\cref{sec:blocking}), so TLOE fails at $\Phi=N$, and whether a finer aggregate is admissible has to be decided from $\Tmin$. A \emph{non-existence result} needs $\Tmin$ itself to be irreversible: this is the case for stage-level controls at the first or last stage of a chain (\cref{cor:stage}) and, numerically, for blocking, where $\Tmin$ turns out to be the full state (\cref{sec:blocking}), and for the autoscaler of \cref{sec:autoscaler}.

\begin{corollary}[Monotonicity in the declared controls]\label{cor:monotone}
Let $\Theta'$ be obtained from $\Theta$ by declaring fewer control directions (fixing the others at their current values), with $\Theta'$ connected and $\theta\in\Theta'$. If an admissible aggregate exists for $\Theta$, one exists for $\Theta'$.
\end{corollary}

\begin{proof}
If $\pi_\theta(x)/\pi_\theta(x')$ does not depend on $\theta\in\Theta$, it does not depend on $\theta\in\Theta'$; so $x\sim_\Theta x'$ implies $x\sim_{\Theta'}x'$, i.e.\ $T_{\Theta'}$ is a function of $\Tmin$. By closure of \cond{R} under coarsening, $\Tmin(X)$ reversible implies $T_{\Theta'}(X)$ reversible; \cref{thm:resolution} concludes.
\end{proof}

Operationally: \textbf{declare the coarsest set of controls you actually act on.} Declaring an extra control can destroy admissibility (\cref{cor:stage}); removing one can restore it, at the price of no longer predicting the response to the removed control.

\subsection{\texorpdfstring{\cond{F} and \cond{R}}{(F) and (R)} are independent}\label{sec:independence}

\begin{proposition}\label{prop:independence}
Neither condition implies the other.
\end{proposition}

\begin{proof}
The counterexamples use the Jackson tandems of \cref{sec:chains} (Poisson arrivals at rate $\lambda$, stations with rates $\mu_k$).

\emph{\cond{R} without \cond{F}.} In the two-stage tandem take $\Theta=\{\ln\lambda\}$ and $\Phi=n_1$. The first station is an M/M/1 queue, so $n_1$ is a stationary birth--death process and is time-reversible. But the conjugate of $\ln\lambda$ is $N-\avg{N}$ with $N=n_1+n_2$ (\cref{prop:conjugate}), and $\E[N-\avg{N}\mid n_1]=n_1-\avg{n_1}$, since $n_1$ and $n_2$ are independent under $\pi$. The part of the conjugate not explained by $n_1$ is therefore $n_2-\avg{n_2}$, which carries the fraction $\Var(n_2)/\Var(N)$ of its variance: $4.444/6.444=0.690$ for $\lambda=1$, $\mu_1=2$, $\mu_2=1.6$. So $n_1$ is not sufficient.

\emph{\cond{F} without \cond{R}.} In a tandem with $K\ge3$ stages take $\Theta=\{\ln\lambda,\ln\mu_1\}$ and $\Phi=(N,n_1)$. By \cref{prop:conjugate}(c), $\Phi$ is the minimal sufficient statistic, a genuine coarse-graining of the state $(n_1,\dots,n_K)$. It is not time-reversible. Under the stationary path law of $\Phi(X)$, a direct jump $(N,n_1)\to(N,n_1-1)$ (a job leaving stage 1) occurs in a unit time interval with positive probability, whereas a direct jump $(N,n_1-1)\to(N,n_1)$ has probability zero, since jobs enter stage 1 only from outside, which raises $N$. Time reversal maps the first event onto the second, so the reversed path law differs from the forward one (\cref{cor:stage}(a)). For $K=2$ the same $\Phi$ is the full state, whose irreversibility is immediate.
\end{proof}

The structural reason is visible in \cref{thm:resolution}: the two conditions move in opposite directions as the resolution changes.

\begin{remark}[Equilibrium is relative to the resolution of the controller]\label{rem:resolution}
\Cref{cor:monotone,cor:stage} make part of this precise: a controller acting on the aggregate load of a chain sees an equilibrium system (\cref{thm:tandem}); a controller that also acts on the first or last stage sees no admissible aggregate, and the direction of the workflow reappears as a failure of reciprocity (\cref{sec:tandem}). The mechanism is already present in the hierarchy of marginal observers of \citet{polettini2019effective}, where bringing some currents to stalling perturbs the others. What this implies for the design of hierarchical control planes is open \hypothesis{} (\cref{sec:limitations}).
\end{remark}

\section{Service chains}\label{sec:chains}

\subsection{The conjugate is dictated by the control}\label{sec:conjugate}

\begin{proposition}[Control-dictated conjugate]\label{prop:conjugate}
In an open Jackson network (\cref{sec:productform}):
\begin{enumerate}[label=(\alph*),leftmargin=*]
\item \emph{Load.} If $\theta=\ln\lambda$ scales all external arrival rates together, then $B=N-\avg{N}$ with $N=\sum_k n_k$, and $N$ is the minimal sufficient statistic: the conjugate is \textbf{global}.
\item \emph{Capacity.} If $\theta=\ln\mu_k$, then $B=-(n_k-\avg{n_k})$: the conjugate is \textbf{local}.
\item \emph{Joint controls.} For $\Theta=\{\ln\lambda,\ln\mu_j\}$ the minimal sufficient statistic is $(N,n_j)$. More generally, a routing weight $\theta$ enters through the traffic equations, and $B=\sum_k(n_k-\avg{n_k})\,\partial_\theta\ln\lambda_k$.
\end{enumerate}
\end{proposition}

\begin{proof}
The traffic equations are linear in the external rates, so every $\lambda_k$ is proportional to $\lambda$ and $\partial\ln\rho_k/\partial\ln\lambda=1$; for a capacity, only $\rho_k$ depends on $\mu_k$, with $\partial\ln\rho_k/\partial\ln\mu_k=-1$, and the normalising term contributes $\rho_k/(1-\rho_k)=\avg{n_k}$. Insert into the score of \cref{sec:productform}. For minimality in (a), $\pi_\theta(n)/\pi_\theta(n')=\lambda^{N(n)-N(n')}\times(\text{a constant})$, which is independent of $\lambda$ if and only if $N(n)=N(n')$; (c) is analogous.
\end{proof}

The formula for the score is that of \citet{comte2025score}; what matters operationally is the pairing between the \emph{type} of control and the \emph{locality} of its conjugate, and the cost of getting it wrong. \Cref{tab:conjugate} shows that cost on the two-stage tandem with $\lambda=1$, $\mu_1=2$, $\mu_2=1.6$. Using the occupancy of stage 2 as the conjugate of a load change, a natural ``local'' choice, overestimates the response of stage 2 about threefold at $t=1$ (0.797 against 0.255; \cref{fig:evidence}a). It recovers the static value only because $n_1$ and $n_2$ are independent under product form, so that $\Cov(n_2,N)=\Var(n_2)$; the error is in the transient. Using the occupancy of stage 1 instead predicts, for the response of stage 2, the wrong sign at all times and a static response of zero. For a load change, the perturbation is carried along the chain by the flow of jobs, and only the global aggregate captures it. For a capacity change, the local conjugate $-n_1$ is exact.

\begin{table}[t]
\centering
\caption{Step responses at $t=1$ (and $t=4$) of the two-stage Jackson tandem: exact values from the perturbed chain, and predictions from \cref{prop:fdt} with the conjugate shown. All values computed exactly on the truncated chain (\cref{app:numerics}). Bold: the conjugate dictated by the control (\cref{prop:conjugate}).}
\label{tab:conjugate}
\resizebox{\textwidth}{!}{%
\begin{tabular}{@{}llrrrrr@{}}
\toprule
Control & Observable & Exact & $B=N$ & $B=n_2$ & $B=n_1$ & $B=-n_1$\\
\midrule
$\delta\ln\lambda$ & $n_2$ & 0.255 (1.416) & \textbf{0.255 (1.416)} & 0.797 (2.164) & $-0.542$ ($-0.748$) & ---\\
$\delta\ln\lambda$ & $N$ & 0.946 (2.918) & \textbf{0.946 (2.918)} & 0.797 (2.164) & 0.149 (0.755) & ---\\
$\delta\ln\mu_1$ & $n_1$ & $-0.691$ ($-1.502$) & --- & --- & --- & \textbf{$-$0.691 ($-$1.502)}\\
$\delta\ln\mu_1$ & $n_2$ & 0.542 (0.748) & --- & --- & --- & \textbf{0.542 (0.748)}\\
\bottomrule
\end{tabular}}
\end{table}

For the capacity control the maximum difference between exact and predicted responses over $t\le8$ is below $2\times10^{-10}$ for both observables.

\subsection{Chains of exponential servers}\label{sec:tandem}

The positive results of this paper rest on a property of chains of exponential single-server stations.

\begin{lemma}\label{lem:tandem}
Consider $K\ge2$ stations in series, each a single exponential server with rate $\mu_k$, unlimited waiting room and FIFO discipline, with service times independent of each other and of the arrival process.
\begin{enumerate}[label=(\roman*),leftmargin=*]
\item \emph{Interchangeability} \citep{weber1979interchangeability,tsoucas1987interchangeability}. Starting from an empty system, for every arrival process independent of the services, the joint law of the arrival and departure counting processes $(A(t),D(t))_{t\ge0}$ does not depend on the order of the stations.
\item \emph{Permutation invariance of total WIP.} If arrivals are Poisson with rate $\lambda<\min_k\mu_k$, the law of the stationary total-WIP process $(N(t))_{t\in\R}$, $N=\sum_kn_k$, does not depend on the order of the stations.
\item \emph{Time-reversibility of total WIP.} Under the hypotheses of (ii), $(N(t))_{t\in\R}$ and $(N(-t))_{t\in\R}$ have the same law.
\end{enumerate}
\end{lemma}

\begin{proof}
(i) Weber's proof establishes the invariance of the departure process for an arbitrary fixed sequence of arrival times, starting from an empty system; Tsoucas and Walrand give a pathwise coupling valid for every arrival process independent of the services. Since the services are independent of the arrivals, the conditional law of $D$ given the arrival path is the law under that fixed path, and integrating over the arrival law gives the joint statement.

(ii) For $t\ge0$, $N(t)=A(t)-D(t)$ for the system started empty, so by (i) the finite-dimensional distributions of $N$ started empty do not depend on the order. The state $X$ of the tandem is an irreducible, positive-recurrent CTMC, so the law of $X(s)$ converges to $\pi$ in total variation as $s\to\infty$, and by the Markov property the finite-dimensional distributions of $(N(s+t_1),\dots,N(s+t_m))$ converge to those of the stationary process. A limit of order-independent quantities is order-independent.

(iii) Let $\tilde X(t)=X(-t)$ be the time reversal of the stationary tandem. It is a stationary Markov chain with rates $\tilde q(x,y)=\pi(y)q(y,x)/\pi(x)$, where $\pi$ is the product-form law \eqref{eq:jackson}. Writing $e_k$ for the $k$-th unit vector, these rates are: $\lambda$ for $x\to x+e_K$ (from $q(x+e_K,x)=\mu_K$ and $\pi(x+e_K)/\pi(x)=\rho_K=\lambda/\mu_K$); $\mu_k$ for $x\to x-e_k+e_{k-1}$, $2\le k\le K$ (the reverse of a forward move from $k-1$ to $k$, which has rate $\mu_{k-1}$, times the ratio $\rho_{k-1}/\rho_k=\mu_k/\mu_{k-1}$); and $\mu_1$ for $x\to x-e_1$ (from $q=\lambda$ and the ratio $1/\rho_1$). So in $\tilde X$ jobs arrive at rate $\lambda$, visit the stations in the order $K,K-1,\dots,1$, and are served at station $k$ at the same rate $\mu_k$ as in the forward chain. Relabelling stations by their position on the reversed route, $\tilde X$ is the forward tandem with rates $(\mu_K,\dots,\mu_1)$. This is the special case of Kelly's theorem on the time reversal of open migration networks \citep[Thm.~2.5]{kelly1979reversibility}. The total WIP of $\tilde X$ is $N(-t)$. By (ii), applied to the reversed order of the stations, its law equals that of $N(t)$.
\end{proof}

We have not found (ii) or (iii) stated in the sources we read, and we claim no novelty for them: both follow in a few lines from (i) and the time reversal. \citet{chao1992reversibility} study the effect of the order of the stations in tandem queues with finite buffers and blocking; their ``reversibility'' is reversal of the order of the stations, not time reversal, and does not contain (ii)--(iii). We checked the fixed-arrival form of (i) exactly for two stations and six fixed arrival times: the joint law of $N$ at four times agrees between the two orders to within $10^{-16}$, while replacing one station by an Erlang-2 server of the same mean breaks the agreement (difference $1.0\times10^{-2}$), so the check can detect a failure (\cref{app:numerics}).

The same argument, applied to blocks of stations, gives a finer statement.

\begin{lemma}[Block interchangeability]\label{lem:block}
Under the hypotheses of \cref{lem:tandem}(ii), fix a station $j$. The law of the stationary process $(N(t),n_j(t))_{t\in\R}$ is invariant under permutations of the rates $(\mu_1,\dots,\mu_{j-1})$ of the stations upstream of $j$, and under permutations of the rates $(\mu_{j+1},\dots,\mu_K)$ of the stations downstream of $j$.
\end{lemma}

\begin{proof}
Start from an empty system and write $A$, $D_{j-1}$, $D_j$ and $D_K$ for the counting processes of external arrivals, of departures from the upstream block into station $j$, of departures from station $j$, and of departures from the chain ($D_{j-1}=A$ if $j=1$). Then $N=A-D_K$ and $n_j=D_{j-1}-D_j$. By \cref{lem:tandem}(i) applied to the upstream block, the joint law of $(A,D_{j-1})$ is invariant under permutations within that block. Given the path of $D_{j-1}$, the process $D_j$ is determined by the services of station $j$, which are independent of the upstream block. Given the path of $D_j$, \cref{lem:tandem}(i) in its fixed-path form makes the conditional law of $D_K$ invariant under permutations within the downstream block, whose services are independent of everything upstream. Hence the joint law of $(A,D_{j-1},D_j,D_K)$ is invariant, and the stationary statement follows as in \cref{lem:tandem}(ii).
\end{proof}

Numerically, the two-time law of $(N,n_3)$ agrees to within $10^{-15}$ between the four-stage chains with rates $(2,1.6,1.8,2.2)$ and $(1.6,2,1.8,2.2)$, and between the five-stage chains $(2,1.6,1.8,2.2,1.7)$ and $(2,1.6,1.8,1.7,2.2)$; moving a rate across station $j$ changes it ($2.2\times10^{-3}$).

\begin{theorem}[TLOE in the tandem]\label{thm:tandem}
For a $K$-station $\cdot$/M/1 tandem with Poisson input and $\lambda<\min_k\mu_k$, and declared controls $\Theta=\{\ln\lambda\}$, the total WIP $N$ is the minimal sufficient statistic, with conjugate $B=N-\avg{N}$, and it is time-reversible. Hence \cond{F} and \cond{R} hold at $\Phi=N$ at every stable throughput, and the tandem is in TLOE in every window over which \cond{S} holds.
\end{theorem}

\begin{proof}
\Cref{prop:conjugate}(a) and \cref{lem:tandem}(iii).
\end{proof}

The proof is two lines given \cref{lem:tandem}; we state the result as a theorem because it is the main positive statement of the paper. It is the precise sense in which a directed system is in equilibrium at the top level. Jobs only move forward, the state $(n_1,\dots,n_K)$ is irreversible and its entropy production at stage resolution is infinite, yet the aggregate fixed by the load control is indistinguishable from its time reversal, and the response of the chain to a change of load follows exactly from the fluctuations of $N$. Near saturation $\taurel$ grows without bound, so \cond{S} eventually fails even though \cond{F} and \cond{R} still hold.

\begin{corollary}[Stage-level controls]\label{cor:stage}
In a $K$-station tandem as in \cref{thm:tandem}, let $\Theta=\{\ln\lambda,\ln\mu_j\}$, so that $\Tmin=(N,n_j)$ (\cref{prop:conjugate}(c)).
\begin{enumerate}[label=(\alph*),leftmargin=*]
\item If $j=1$ or $j=K$, $\Tmin(X)$ is not time-reversible, and no coarse-graining satisfies \cond{F} and \cond{R}.
\item If $K$ is odd, $j=(K+1)/2$, and the multiset of upstream rates $\{\mu_1,\dots,\mu_{j-1}\}$ equals the multiset of downstream rates $\{\mu_{j+1},\dots,\mu_K\}$, then $\Tmin(X)$ is time-reversible, and an admissible aggregate exists ($\Tmin$ itself) for every value of the controls.
\item In every other case we computed, $\Tmin(X)$ is not time-reversible \numerical; whether (b) is also necessary in general is open.
\end{enumerate}
\end{corollary}

\begin{proof}
(a) Take $j=1$. Under the stationary path law of $\Tmin(X)$, the event ``a direct jump $(N,n_1)\to(N,n_1-1)$ occurs in $[0,1]$'' (a job leaving stage 1 for stage 2) has positive probability, because its stationary intensity $\sum\pi(x)\mu_1$ over the states with $n_1\ge1$ and $N$ fixed is positive. The event ``a direct jump $(N,n_1-1)\to(N,n_1)$ occurs in $[0,1]$'' has probability zero, since jobs enter stage 1 only from outside, which raises $N$. Time reversal maps the first event onto the second, so the reversed path law differs from the forward one. For $j=K$, the same argument applies to the jumps $(N,n_K)\to(N,n_K+1)$ (a job entering stage $K$ from stage $K-1$), which have positive intensity, and their reverses, which have none. \Cref{thm:resolution} concludes.

(b) By the proof of \cref{lem:tandem}(iii), the time reversal of the stationary tandem is the forward tandem with rates $(\mu_K,\dots,\mu_1)$, in which station $j$ sits at position $K+1-j=j$. So $(N(-t),n_j(-t))_t$ has the law of $(N,n_j)$ in a chain whose upstream block has rates $(\mu_K,\dots,\mu_{j+1})$ and whose downstream block has rates $(\mu_{j-1},\dots,\mu_1)$. If the two multisets coincide, these are permutations of the original upstream and downstream rates, and by \cref{lem:block} the law is that of the original $(N,n_j)$. The hypothesis involves neither $\lambda$ nor $\mu_j$, so it holds on the whole control family.
\end{proof}

\begin{table}[t]
\centering
\caption{Largest asymmetry $\max|J-J\T|$ of the two-time joint law $J$ of $(N,n_j)$ at lags 0.5, 1 and 2, exact computation on chains truncated at a total of 11 to 24 jobs (values stable under the truncation; \cref{app:numerics}).}
\label{tab:stage}
\small
\begin{tabular}{@{}clllr@{}}
\toprule
$K$ & rates $(\mu_1,\dots,\mu_K)$ & statistic & condition of \cref{cor:stage}(b) & max asymmetry\\
\midrule
3 & $(2,1.6,2)$ & $(N,n_2)$ & holds & $\le6\times10^{-16}$\\
3 & $(2,2,2)$ & $(N,n_2)$ & holds & $\le3\times10^{-16}$\\
5 & $(2,1.6,1.8,2,1.6)$ & $(N,n_3)$ & holds (not mirror-symmetric) & $\le7\times10^{-16}$\\
5 & $(1.6,2,1.8,2,1.6)$ & $(N,n_3)$ & holds & $\le9\times10^{-16}$\\
\midrule
3 & $(2,1.6,2.05)$ & $(N,n_2)$ & fails & $1.5\times10^{-4}$\\
3 & $(2,1.6,1.8)$ & $(N,n_2)$ & fails & $5.5\times10^{-4}$\\
3 & $(3,1.3,2)$ & $(N,n_2)$ & fails & $2.1\times10^{-3}$\\
5 & $(2,1.6,1.8,2,1.7)$ & $(N,n_3)$ & fails & $7.6\times10^{-5}$\\
4 & $(2,1.6,1.6,2)$ & $(N,n_2)$, $(N,n_3)$ & fails (no middle stage) & $2.5\times10^{-3}$\\
3 & $(2,1.6,2)$ & $(N,n_1)$, $(N,n_3)$ & case (a) & $1.6\times10^{-2}$\\
\bottomrule
\end{tabular}
\end{table}

The middle stage is special because it is the only station that time reversal maps to itself (\cref{tab:stage}). In the two-stage tandem the failure under $\Theta=\{\ln\lambda,\ln\mu_1\}$ is also visible as a failure of reciprocity (\cref{prop:consequences}(c)): the step response of $N$ to $\delta\ln\mu_1$ is $-0.149$ at $t=1$ ($-0.755$ at $t=4$), while that of $-n_1$ to $\delta\ln\lambda$ is $-0.691$ ($-1.502$). The value 0.149 also appears in \cref{tab:conjugate}, as the response of $N$ to $\delta\ln\lambda$ predicted with $B=n_1$. This is not a coincidence: both equal $C_{N n_1}(0)-C_{N n_1}(1)=\Var(n_1)-C_{N n_1}(1)=2-1.851$, up to the sign, since the conjugate of $\ln\mu_1$ is $-n_1$.

\subsection{Blocking breaks both conditions, weakly}\label{sec:blocking}

Product form is lost when a stage has a finite buffer. We model \emph{blocking after service}: a job that finishes at stage 1 while stage 2 is full waits at stage 1, and stage 1 resumes when space appears. Other disciplines (blocking before service, communication blocking) have different stationary laws, and the numbers below are for blocking after service only. \Cref{tab:blocking} compares the Jackson tandem (model J) with the same tandem and a buffer of 3 jobs at stage 2 (model B), with $Y=N$ and $\Theta=\{\ln\lambda\}$. Both are solved exactly on a chain truncated at 70 jobs in stage 1 (in model B, $P(n_1=70)=9\times10^{-12}$; in model J, stage 2 is also truncated at 70, with $P(n_2=70)\approx2\times10^{-15}$).

\begin{table}[t]
\centering
\caption{Exact tests on models J and B (\cref{app:numerics}). Values at machine precision are given as bounds.}
\label{tab:blocking}
\small
\begin{tabularx}{\textwidth}{@{}Xll@{}}
\toprule
Quantity & Model J & Model B\\
\midrule
Deviation from product form, max over $n\le5$ of the non-additive part of $\ln\pi$ & $\le10^{-14}$ & 1.8\\
\cond{F}: fraction of $\Var(B)$ not explained by $N$, $\Var(B-\E[B\mid N])/\Var(B)$ & $\le10^{-15}$ & $1.3\times10^{-2}$\\
\cond{R}: two-time asymmetry $\E[N_0^2N_t]-\E[N_0N_t^2]$ at $t=4$ & $\le10^{-13}$ & 0.47\\
\cond{R}: lagged-moment asymmetry $\|K-K\T\|/\|K\|$, 13-bin indicator dictionary on $N$, $\tau=0.5$ & $\le10^{-14}$ & $9.2\times10^{-3}$\\
\cond{R}: three-time asymmetry of $N$ over 8 random triples $f,g,h$ (lag 0.7); positive control on full-state functions in brackets & $\le10^{-15}$ ($2.3\times10^{-2}$) & $9.3\times10^{-3}$ ($2.0\times10^{-2}$)\\
Response of $N$ to $\delta\ln\lambda$ at $t=1$: exact vs prediction with $B=N$ & 0.946 vs 0.946 & 0.952 vs 0.922\\
Same, with the exact conjugate $\partial\ln\pi$ (full state), max error over $t\le8$ & $\le10^{-9}$ & $\le10^{-9}$\\
\bottomrule
\end{tabularx}
\end{table}

In model J the values are consequences of \cref{thm:tandem} and serve as a check of the numerical method; the stage-permutation invariance of the two-time law of $N$ is reproduced to within $10^{-15}$. In model B both conditions fail, but the violations are small: $N$ leaves 1.3\% of the variance of the conjugate unexplained, the normalised lagged-moment asymmetry is below 1\%, and the prediction with $B=N$ is off by 3.2\% at $t=1$, 7.7\% at $t=4$ and 11.8\% in the static limit. The general identity still holds with the exact conjugate, which is no longer a function of $N$. The top-level symmetry of $N$ in model J is therefore not an artefact of aggregation: it disappears when the system leaves the product-form regime. In model B the generalised eigenvalues of the lagged moments on this dictionary remain real although $K$ is asymmetric, so complex eigenvalues are a one-sided signal; the asymmetry of $K$ is the better statistic.

\paragraph{No admissible aggregate, and how the violation vanishes \numerical.} Under the load knob, no two states of model B with $n_1\le12$ keep a fixed probability ratio when $\lambda$ ranges over $\{0.8,0.9,1.0,1.1\}$: the smallest difference between the changes of their log-probabilities is $3.4\times10^{-3}$. On that range the minimal sufficient statistic is therefore the full state $(n_1,n_2)$, whose direct jumps $(n_1,n_2)\to(n_1-1,n_2+1)$ have no reverse, so by \cref{thm:resolution} no aggregate is admissible: blocking is a non-existence result, not only a failure at $\Phi=N$. The size of the violation at $N$ nonetheless falls quickly with the buffer. With buffers of 2, 3, 5, 8 and 12 jobs at stage 2, the stage is full 31\%, 17\%, 5.9\%, 1.4\% and 0.2\% of the time, and the prediction with $B=N$ underestimates the static response by 18\%, 12\%, 4.0\%, 0.6\% and 0.04\%. Exact TLOE does not return with a finite buffer, but buffer sizing can make the violation negligible: configuration is a design lever on how far a system is from TLOE, distinct from the declared controls.

\subsection{A strong counter-model: a hysteretic autoscaler}\label{sec:autoscaler}

Blocking is a mild violation. A common mechanism in elastic services produces a strong one. Model A is a queue with Poisson arrivals at rate $\lambda=1.2$, served by $r\in\{1,2\}$ replicas of rate $\mu=1$ each (service rate $\min(n,r)\mu$). An autoscaler adds the second replica at rate $\alpha$ when $r=1$ and $n\ge8$, and removes it at rate $\alpha$ when $r=2$ and $n\le1$, so $1/\alpha$ is its reaction time and the gap between the thresholds is its hysteresis. With one replica the queue is unstable at this load ($\lambda>\mu$), so the autoscaler is needed. The state is $(n,r)$, the top-level observable is $Y=N=n$, and the control is $\ln\lambda$.

\begin{table}[t]
\centering
\caption{Model A, exact computation (truncation at $n=160$), for several reaction times $1/\alpha$, compared with models J, B (buffer 3) and B2 (buffer 2). \cond{F}: fraction of the conjugate's variance not explained by $N$. \cond{R}: normalised asymmetry $\|K-K\T\|/\|K\|$ and largest standardised antisymmetric component $|J_{ab}-J_{ba}|/\sqrt{p_ap_b}$ of the 13-bin indicator dictionary at lag $\max(0.5,\tauint/4)$. Prediction errors: supremum over $t\le3\tauint$ of the error of the passive prediction of the response of $N$ to $\delta\ln\lambda$, relative to the static response, with the conjugate $B=N$ of \cref{prop:conjugate} and with the top-level conjugate $B^Y$ of \cref{sec:testF}.}
\label{tab:autoscaler}
\resizebox{\textwidth}{!}{%
\begin{tabular}{@{}lrrrrrrrr@{}}
\toprule
Model & $\alpha$ & $\tauint$ & $\E[N]$ & \cond{F} unexpl. & $\frac{\|K-K\T\|}{\|K\|}$ & max std.\ asym. & error, $B=N$ & error, $B^Y$\\
\midrule
J & --- & 7.6 & 2.67 & 0 & 0 & 0 & 0 & 0\\
B (buffer 3) & --- & 17.0 & 3.19 & 0.013 & 0.027 & 0.010 & 0.11 & 0.008\\
B2 (buffer 2) & --- & 51.4 & 4.87 & 0.005 & 0.017 & 0.005 & 0.17 & 0.004\\
\midrule
A & 0.003 & 331.9 & 27.60 & 0.43 & 0.036 & 0.036 & 22.3 & 0.40\\
A & 0.01 & 135.5 & 13.29 & 0.51 & 0.065 & 0.037 & 8.04 & 0.35\\
A & 0.03 & 43.3 & 7.59 & 0.62 & 0.053 & 0.023 & 4.01 & 0.46\\
A & 0.1 & 11.8 & 5.32 & 0.62 & 0.053 & 0.020 & 1.90 & 0.45\\
A & 0.3 & 5.3 & 4.52 & 0.48 & 0.056 & 0.022 & 0.93 & 0.28\\
A & 1 & 4.3 & 4.14 & 0.32 & 0.062 & 0.026 & 0.61 & 0.16\\
A & 10 & 4.4 & 3.93 & 0.21 & 0.070 & 0.028 & 0.44 & 0.10\\
A & 100 & 4.4 & 3.90 & 0.19 & 0.071 & 0.029 & 0.41 & 0.09\\
\bottomrule
\end{tabular}}
\end{table}

\paragraph{No admissible aggregate exists for model A at the reaction times computed \numerical.} The argument does not need the minimal sufficient statistic. Let $B$ be the exact score of $\ln\lambda$ at the operating point. Any sufficient $\Phi$ satisfies $B=b(\Phi)$ for some $b$, so the process $B(X)$ is a coarsening of $\Phi(X)$; by closure of \cond{R} under coarsening (\cref{sec:resolution}), if $\Phi(X)$ were time-reversible, so would be $B(X)$. A time-reversible coarse-grained process satisfies \emph{lumped detailed balance}: for every pair of distinct classes $C_a$, $C_b$ of the partition of $S$ by the level sets of $B$,
\begin{equation}\label{eq:lumpedDB}
j_{ab}:=\sum_{x\in C_a,\;y\in C_b}\pi(x)\,q(x,y)\;=\;j_{ba},
\end{equation}
because $j_{ab}$ is the stationary rate of direct jumps of $B(X)$ from $C_a$ to $C_b$, and time reversal maps those jumps onto jumps from $C_b$ to $C_a$. In model A, \eqref{eq:lumpedDB} fails at every reaction time computed ($\alpha\in\{0.003,0.03,0.1,1,100\}$), on chains truncated at $n=160$ and at $n=240$. The largest imbalance is always between two neighbouring one-replica states, for example $(2,1)$ and $(3,1)$ at $\alpha=0.1$, with fluxes $6.83\times10^{-2}$ and $5.40\times10^{-2}$; across $\alpha$ the imbalance is 17--32\% of the larger flux. It is the net probability current around the hysteresis loop (the queue grows with one replica, a replica is added at $n\ge8$, the queue drains with two, a replica is removed at $n\le1$), which flows through the one-replica states in the direction of increasing $n$. Each of the two states is a level set of $B$ on its own, and its score differs from that of every other state by at least $0.048$, far above the numerical error of $B$, so the imbalance is not an artefact of discretising the level sets. Hence $B(X)$ is not time-reversible, and no coarse-graining satisfies \cond{F} and \cond{R}. As a cross-check, the minimal sufficient statistic appears to be the full state: for every pair of states with $n\le20$, the ratio of their stationary probabilities changes with $\lambda$ (the closest pair, at $\alpha=100$, is $(0,2)$ and $(8,1)$, two rare states whose log-ratio changes by $1.4\times10^{-7}$ under a 5\% change of $\lambda$); this was checked at $\alpha\in\{0.003,0.1,100\}$ and is not needed by the argument. The result is numerical: it covers the reaction times and truncations listed, not every $\alpha>0$.

Three observations follow.
\begin{itemize}[leftmargin=*]
\item \textbf{The violation of \cond{F} is large.} $N$ leaves 19--62\% of the conjugate's variance unexplained, against 1.3\% for blocking, and the passive prediction with $B=N$ is wrong by 41\% to 190\% of the static response for reaction times up to the relaxation time, and by far more for slower autoscalers, whose long cycles dominate the variance of $N$. Even the top-level conjugate $B^Y$, which reproduces the static response exactly (\cref{prop:identifiability}), misses the transient by 9--46\%.
\item \textbf{A fast autoscaler is not averaged out.} The violation is smallest but does not vanish as $\alpha$ grows (19\% at $\alpha=100$). The hysteresis keeps the replica count a hidden variable whose memory lives on the time scale of the queue itself, whatever the reaction delay, so the regime $\tau_{\mathrm{mech}}\ll\taurel$ of \cref{sec:scales} does not apply: the relevant time scale is the dwell time in the hysteresis band, not $1/\alpha$.
\item \textbf{The time asymmetry of $N$ is moderate.} The normalised asymmetry of the lagged moments is 4--7\%, 1.3 to 2.6 times that of blocking at the same lag convention, and the largest standardised component is 0.02--0.04, two to four times that of blocking. The autoscaler shows up mainly as a failure of \cond{F}, and only moderately as a failure of \cond{R} at $\Phi=N$; \cref{sec:finite} measures what this means for detection.
\end{itemize}

\section{Testing the conditions on telemetry}\label{sec:tests}

Each condition of \cref{def:tloe} has a test that runs on a window of top-level telemetry. None needs the parameters of a queueing model: the Jackson network is the \emph{reference model} on which results are proved and falsified, not a model the operator must calibrate. This section specifies the tests; \cref{sec:finite} evaluates the prediction and the reversibility test on simulated telemetry and estimates the cost of the sufficiency test, which we do not run.

\subsection{T-S: stationarity and time-scale separation}\label{sec:testS}

The four inequalities of \cond{S} are not equally testable (\cref{rem:S}); \cref{tab:testS} summarises their status.

\begin{table}[t]
\centering
\caption{Status of the parts of \cond{S} and how each is handled.}
\label{tab:testS}
\small
\begin{tabularx}{\textwidth}{@{}llX@{}}
\toprule
Part of \cond{S} & Status & Procedure\\
\midrule
Law of $Y$ stationary over $\Tobs$ & tested & after discarding an initial transient (e.g.\ MSER, \citealp{franklin2008stationarity}), compare the marginal law and autocovariance of $Y$ across sub-windows with a two-sample test and a multiplicity correction, or apply a stationarity or change-point test such as KPSS \citep{kwiatkowski1992testing} or CUSUM \citep{page1954continuous}\\
$\taurel\ll\Tobs$ & tested & estimate $\taurel$ (implied time scales, \cref{sec:testR}) and $\tauint$ from the window; require $\Tobs/\tauint$ above the values of \cref{sec:finite}\\
$\Tobs\ll\taudrift$ & monitored & track the window statistics of $Y$ across consecutive windows; drift shows up as a trend in them\\
$\taulocal\ll\taurel$ & assumed & not observable in telemetry sampled more slowly than $\taulocal$; it states what the operator chooses to integrate out (\cref{sec:scales})\\
\bottomrule
\end{tabularx}
\end{table}

Two cautions apply. MSER and similar methods locate where a steady state starts, assuming one exists; they do not test existence, and steady performance may not exist at all \citep{antoch2026experimental}. And failing to reject stationarity in one window is not evidence of stationarity. \textbf{Self-excited load} is a typical way \cond{S} fails: when retries are modelled as a Hawkes process with branching ratio $n<1$, the mean rate is $\lambda_0/(1-n)$ and diverges as $n\to1$ \citep{hawkes1971spectra}, and $\taurel$ grows without bound. Loss of time-scale separation is itself a ``not OK'' signal. Growth of variance and correlation time near saturation (critical slowing down) is expected on general grounds, but generic early-warning indicators have performed poorly in realistic network simulations, where a simple threshold did better \citep{dabrowski2018evaluating}; we do not rely on them.

\subsection{T-F: sufficiency, and what top-level data cannot see}\label{sec:testF}

Testing \cond{F} requires care, because part of what it asserts is invisible in top-level data. Let $\pi^Y_\theta$ be the stationary law of $Y$ and $B^Y_i(y)=\partial_{\theta_i}\ln\pi^Y_\theta(y)$ its score, which has mean zero.

\begin{proposition}[Identifiability on top-level data]\label{prop:identifiability}
\leavevmode
\begin{enumerate}[label=(\alph*),leftmargin=*]
\item $B^Y_i(y)=\E[\partial_{\theta_i}\ln\pi_\theta(X)\mid Y=y]$. Consequently $\partial_{\theta_i}\avg{a(Y)}=\Cov(a(Y),B^Y_i(Y))$ for every function $a$, \textbf{whether or not \cond{F} holds}.
\item Under \cond{F}, $B^Y_i=B_i-\avg{B_i}$, and the dynamic identity $R_{A,i}(\tau)=C_{A,B^Y_i}(0)-C_{A,B^Y_i}(\tau)$ holds for all $A=a(Y)$ and $\tau\ge0$. Without \cond{F}, it still holds at $\tau=0$ and as $\tau\to\infty$, and can fail only at intermediate $\tau$.
\end{enumerate}
\end{proposition}

\begin{proof}
(a) is the missing-information identity: the score of the observed part is the conditional expectation of the full score \citep[Eq.~2.13]{orchard1972missing}; the static statement is the score identity for the law of $Y$. (b) The exact response is $C_{A,B}(0)-C_{A,B}(\tau)$ with the full conjugate $B=B(X)$, and $C_{A,B}(\tau)=\E\big[\E[a(Y_\tau)\mid X_0]\,B(X_0)\big]-\avg{A}\avg{B}$. Replacing $B(X_0)$ by $\E[B(X_0)\mid Y_0]$ leaves this unchanged when $\E[a(Y_\tau)\mid X_0]$ is a function of $Y_0$, which is the case at $\tau=0$ and in the limit $\tau\to\infty$, but not in general in between.
\end{proof}

\begin{remark}[Size of the transient error]\label{rem:transientbound}
Write $B=B^Y+\tilde B$ for a control $i$, with $\tilde B=B-\E[B\mid Y]$ the part of the conjugate that is invisible from $Y$. For $A=a(Y)$, the error of the prediction made with $B^Y$ is $-\Cov\big(a(Y_\tau),\tilde B(X_0)\big)$, and
\[
\big|\Cov\big(a(Y_\tau),\tilde B(X_0)\big)\big|\;\le\;\sqrt{\Var\tilde B}\;\big\|\E[a(Y_\tau)\mid X_0]-\E[a(Y_\tau)\mid Y_0]\big\|_{L^2(\pi)}\;\le\;\sqrt{\Var\tilde B}\,\sqrt{\Var a(Y)}.
\]
Indeed, with $g(X_0)=\E[a(Y_\tau)\mid X_0]$ and $h(Y_0)=\E[g(X_0)\mid Y_0]$, the covariance equals $\E[g\,\tilde B]=\E[(g-h)\,\tilde B]$, because $\tilde B$ has mean zero and is orthogonal to every function of $Y_0$; Cauchy--Schwarz gives the first inequality, and $\|g-h\|^2=\Var g-\Var h\le\Var g\le\Var a(Y)$ the second. The middle factor is zero at $\tau=0$ and tends to zero as $\tau\to\infty$, so the error is a bump between the two limits, bounded by the product of how much of the conjugate is hidden and how much the hidden state predicts the future of $A$ beyond what $Y_0$ already does. The quantity that the dynamic test of \cond{F} with the estimated conjugate has to detect therefore also bounds the error of the prediction made with that conjugate: a violation too weak to detect also biases the prediction with $B^Y$ little (in model B, at most 0.8\% of the static response; \cref{tab:autoscaler}). The larger risk is a hypothesised conjugate, whose error does not vanish in the static limit (11.8\% for $B=N$ in model B).
\end{remark}

There are therefore two different tests of \cond{F}, and they must not be confused.
\begin{enumerate}[leftmargin=*]
\item \textbf{T-F with a hypothesised conjugate (a joint test).} Fix the conjugate from a model, e.g.\ $B=N$ for a change of load (\cref{prop:conjugate}), and compare its predictions with the system's response. A static comparison already has power: in model B, $\Var(N)=9.78$ against the exact static response 11.10. But such a test checks \cond{F} \emph{jointly} with the hypothesis about the conjugate, not sufficiency alone; it is the form the KL test of \cref{sec:kl} takes.
\item \textbf{T-F with the estimated top-level conjugate (the sufficiency test).} Estimate $B^Y$ from the laws of $Y$ in two windows at nearby control settings, $\hat B=\delta^{-1}\ln(\hat\pi^Y_{\theta+\delta}/\hat\pi^Y_\theta)$ (a density-ratio estimate, \citealp{sugiyama2012density}). By \cref{prop:identifiability}(a), $\hat B$ reproduces every static sensitivity of every top-level observable whether or not $Y$ is sufficient, so \textbf{a static test with the estimated conjugate has no power}; verifying a static FDT does not establish equilibrium \citep{owen2020universal}. Only the time-resolved response of $Y$ to a change of the control, produced by the routine variation of the controls between windows, can discriminate. In model B the effect is real but small: with $B^Y$, the predicted response of $N$ to $\delta\ln\lambda$ is 1.018 at $t=1$ against 0.952 exact (7\%), 3.149 against 3.067 at $t=4$ (3\%), 7.188 against 7.151 at $t=16$ (0.5\%), and the static value is exact (11.095). In model A it is large: 9--46\% (\cref{tab:autoscaler}).
\end{enumerate}

The second test is the one that tests sufficiency, and its cost is much higher than that of the first (\cref{sec:tfcost}). \Cref{prop:identifiability} has a counterpart in system identification, where a model with too few states can be fitted to reproduce the static gain of a plant while missing its transient \citep{ljung1999system}; here the reduced description is the top-level aggregate, and the transient is the only place the missing information shows.

\subsection{T-R: time-reversibility}\label{sec:testR}

\begin{itemize}[leftmargin=*]
\item \textbf{Statistic.} The antisymmetric part of lagged two-time statistics of $Y$ on a dictionary of functions,
\[
D_{ab}(\tau)=\E[\psi_a(Y_0)\psi_b(Y_\tau)]-\E[\psi_b(Y_0)\psi_a(Y_\tau)],
\]
at a few lags. The dictionary must be \emph{nonlinear} (indicators of ranges of $Y$, powers) or involve several observables: the autocovariance of a single observable is always symmetric, and a stationary scalar Gaussian process is always reversible \citep{weiss1975time}, so second-order statistics of one linear observable cannot reveal irreversibility. This is the cross-moment family of time-reversibility tests \citep{ramsey1996time}, reviewed by \citet{goto2026modelfree}; for discretised $Y$ it is close to the transition-count tests for Markov chains \citep{edwards1979large,steuber2012testing}. Finite-dictionary tests check necessary conditions only.
\item \textbf{Null distribution.} A block bootstrap of the series, centred at the observed statistic, gives a null that needs no model and no Markov assumption (\cref{sec:finite-design} gives the implementation, \cref{sec:finite-tr} its size and power). Moving-block, stationary and automatically tuned block bootstraps are alternatives \citep{politis1994stationary,politis2004automatic}. With finite data, \cond{R} is only ever \emph{not rejected}.
\item \textbf{Validity of a rejection.} A population-level asymmetry of $Y$ is a valid violation verdict without any Markov assumption, because coarse-graining cannot create irreversibility in the exact law (\cref{sec:consequences}). What can create spurious asymmetry is (i) finite samples, handled by the null, and (ii) measurement artefacts, above all clock skew or non-synchronised sampling when $Y$ aggregates counters from several nodes. Discretising $Y$ is safe; unsynchronised timestamps are not.
\item \textbf{Monitoring.} The test controls the false-alarm rate \emph{per window}. A monitoring system that tests every window, especially overlapping ones, needs a sequential correction, for instance a CUSUM on the statistic or a level spent across looks \citep{page1954continuous,wald1945sequential}.
\item \textbf{Spectral use.} Reading relaxation times from the generalised eigenvalues of the pencil $(K(\tau),K(0))$ (\cref{prop:consequences}(d)--(e)) requires, in addition, that $Y$ be approximately Markov at the lags used. Independence of the implied time scales $t_k=-\tau/\ln\kappa_k(\tau)$ from $\tau$ \citep{prinz2011markov} is a necessary check only: in the tandem it passes although $N$ is not Markov (\cref{sec:consequences}). Transfer-operator estimators such as TICA impose the symmetry that T-R tests \citep{perezhernandez2013identification}, so the test must be run on the unsymmetrised estimate.
\end{itemize}

\subsection{Where the parameters come from}\label{sec:parameters}

TLOE needs no queueing parameters: T-S uses the statistics of $Y$, T-F the conjugate (estimated from two windows) and the measured transient, T-R the lagged statistics of $Y$, and the prediction the two-time covariances of $Y$. Two further uses of the framework do involve parameters.
\begin{itemize}[leftmargin=*]
\item \textbf{Discovering the top-level variables} \hypothesis. Regressing $\hat B$ on candidate observables (total requests in flight per chain, per-stage queues, pool occupancies) and keeping the smallest set that explains it estimates the minimal sufficient statistic. With \cref{thm:resolution}, this turns the choice of $\Phi$ into an estimation problem.
\item \textbf{Explicit models}, for what-if analysis beyond linear response or to design $\Phi$ in advance, come from software performance engineering rather than network modelling: arrival rates from ingress logs, routing from call-graph frequencies in distributed traces, service demands from span durations or regression on the utilisation law \citep{spinner2015evaluating}, software contention through layered queueing networks \citep{woodside1995stochastic}, and non-exponential service through phase-type fits, which keep the state Markov. Little's law gives a per-component consistency check. If T-F fails at $N$, product form fails, whatever the fitted parameters say.
\end{itemize}

\subsection{Predictive consistency as a test of \texorpdfstring{\cond{F}}{(F)}}\label{sec:kl}

\citet{rouska2025equilibrium} measure the distance from equilibrium as a KL divergence between the behaviour expected by an internal model and the behaviour observed after a perturbation. TLOE supplies the expected distribution without a model. Under \cond{F}, a small control change $\delta\theta$ acts on the top-level law as an exponential tilt by the conjugate,
\[
\pi^Y_{\theta+\delta\theta}(y)\propto\pi^Y_\theta(y)\,\exp\!\big(\delta\theta\cdot B(y)\big)\,\big(1+O(|\delta\theta|^2)\big),
\]
so the ``expected'' distribution after a control change is the tilt of the \emph{unperturbed} top-level law by the control-dictated conjugate.

\begin{proposition}[KL bridge]\label{prop:kl}
Let $\pi_{\mathrm{exp}}\propto\pi^Y_\theta\exp(\delta\tilde B)$ be the tilt by a hypothesised conjugate $\tilde B$ and $\pi_{\mathrm{obs}}=\pi^Y_{\theta+\delta}$ the observed law. Then
\[
\KL(\pi_{\mathrm{obs}}\,\|\,\pi_{\mathrm{exp}})=\tfrac12\,\delta^2\Var\big(B^Y-\tilde B\big)+O(\delta^3).
\]
It vanishes to second order if and only if $\tilde B=B^Y+\text{const}$, $\pi^Y$-almost surely.
\end{proposition}

\begin{proof}
$\ln\pi_{\mathrm{obs}}-\ln\pi_{\mathrm{exp}}=\delta(B^Y-\tilde B)+c(\delta)+O(\delta^2)$ with a normalising constant $c$; expand the divergence to second order.
\end{proof}

With $\tilde B=N$, the KL therefore tests whether $N$ is the conjugate, i.e.\ \cond{F} jointly with \cref{prop:conjugate}; with $\tilde B=B^Y$ it is uninformative, by \cref{prop:identifiability}. In a tandem of $\cdot$/M/1 stations the tilt by $N$ is exact for every $\delta$, not only to second order: summing the product form over the states with a given $N$ gives $\pi^N_\lambda(N)=C(\lambda)\,\lambda^N h(N)$ with $h(N)=\sum_{n_1+\dots+n_K=N}\prod_k\mu_k^{-n_k}$, which does not depend on $\lambda$, so the law of $N$ is an exponential family in $\ln\lambda$ for any $K$ and any rates. Accordingly, in model J the divergence is below $3.2\times10^{-16}$ for $\delta\in\{0.05,0.1,\pm0.2\}$. In model B, $\tfrac12\Var(B^Y-N)=0.0930$ and the ratio $\KL/\delta^2$ is 0.0940, 0.1030, 0.159 and 0.292 at $\delta=10^{-3}$, $10^{-2}$, $5\times10^{-2}$ and $10^{-1}$; the first two give a linear correction of slope 1.00 and an extrapolated coefficient of 0.09297, against the predicted 0.09303.

Condition \cond{R} is also a KL: the relative-entropy rate between $Y$-paths and their time reversals. TLOE can therefore be stated as \textbf{stationarity plus two vanishing KL divergences}, the language of predictive consistency, now with an exact identity underneath. The comparison also shows why agreement between a model and the system is not the same as equilibrium: a perfect model of model B gives a divergence of zero, although B is not in TLOE; and a system in TLOE observed through a poor model gives a positive one.

\subsection{Bayesian networks and the Markov blanket of the controls}\label{sec:dbn}

A dynamic Bayesian network (DBN) learned on top-level telemetry is a factored Markov model of $Y$, and a natural vehicle for the tests, provided it tests the \emph{system} rather than the \emph{model}. That needs two separate steps. First, validate the DBN on unperturbed data (held-out predictive likelihood, Markov order). Second, read the conditions off the validated DBN: \cond{S} as stability of its conditional probability tables across windows; \cond{F} by comparing the observed post-change law with the tilt of the pre-change stationary law (\cref{sec:kl}), not with the DBN re-run under the change; and \cond{R} by comparing the learned two-slice joint $J(y_t,y_{t+\Delta t})$ with its transpose. The quantity $(\Delta t)^{-1}\sum J\ln(J/J\T)$ is the entropy-production rate per unit time of the fitted chain, so this test is also a KL; its standard nulls assume that the discretised $Y$ is Markov \citep{edwards1979large,steuber2012testing}, which \cref{sec:consequences} shows need not hold.

\begin{proposition}[Markov blanket]\label{prop:blanket}
Suppose the stationary law of $X$ factorises as a Bayesian network in which the controls $\theta$ enter as root nodes. Then the variables of the Markov blanket of $\theta$, its children together with their other parents, form a sufficient statistic for $\theta$.
\end{proposition}

\begin{proof}
$\pi_\theta(x)=\prod_{i\notin\mathrm{ch}(\theta)}p(x_i\mid\mathrm{pa}_i)\prod_{i\in\mathrm{ch}(\theta)}p(x_i\mid\mathrm{pa}_i,\theta)$; the second product depends on $x$ only through the blanket. Apply Fisher--Neyman.
\end{proof}

Placing parameters as parent nodes of a Bayesian network is standard \citep{spiegelhalter1993bayesian}, and the proposition is a two-line consequence of the factorisation; we state it because it gives a precise meaning to the Markov-blanket view of computing-continuum systems \citep{dustdar2023distributed}: the Markov blanket of the controls is a sufficient statistic, hence a refinement of the minimal top-level statistic $\Tmin$, and so is every aggregate that satisfies \cond{F}. Two caveats apply. The blanket is \emph{a} sufficient statistic, not necessarily the minimal one, and an admissible aggregate need not refine it: in the tandem the blanket contains $(n_1,n_2)$, while the minimal statistic for the load is $N$. And it must be read off the graph of the \emph{stationary} law, which is generally denser than the two-slice graph of the DBN \citep{boyen1998tractable}; in practice, test it with $\hat B$ (\cref{sec:parameters}).

\section{How much telemetry the prediction and the tests need}\label{sec:finite}

All results so far are exact: they hold for the stationary law and for infinitely long observation. An operator has a finite window. This section asks two questions on simulated telemetry of the reference models:
\begin{itemize}[leftmargin=*]
\item \textbf{(H1) Prediction.} When TLOE holds, how accurate is the passive prediction of the response to a control change, as a function of the window length?
\item \textbf{(H2) Detection.} When a condition fails, how often does a test on top-level telemetry detect it at a 5\% false-alarm rate per window, as a function of the window length?
\end{itemize}
H2 is evaluated for the reversibility test T-R, which we implement. For the sufficiency test T-F we only estimate the cost (\cref{sec:tfcost}). All results of this section are numerical.

\subsection{Design}\label{sec:finite-design}

\begin{itemize}[leftmargin=*]
\item \textbf{Models.} J (Jackson tandem), B3 (buffer of 3 at stage 2, model B of \cref{sec:blocking}), B2 (buffer of 2) and A (the autoscaler of \cref{sec:autoscaler} with $\alpha=0.1$). The tandems have $\lambda=1$, $\mu_1=2$, $\mu_2=1.6$. In the tandems the violations are weak; in A they are strong (\cref{tab:autoscaler}).
\item \textbf{Telemetry.} The total number of jobs $N$, sampled every $\Delta=0.5$ time units, as a monitoring system would; time is measured in units of the mean inter-arrival time of the tandem ($\lambda=1$).
\item \textbf{Window unit.} Window lengths are $\Tobs=k\,\tauint$ with $k\in\{10,30,100,300,1000,3000,10000\}$, where $\tauint$ is the integrated autocorrelation time of $N$, computed exactly: J 7.6, B3 17.0, B2 51.4, A 11.8. $\tauint$ sets the effective sample size of a window; the slowest relaxation time $\taurel$ is larger (for J, 14.2 against $\tauint=7.6$; \cref{sec:state}), so a window of $k\,\tauint$ is about $0.5\,k\,\taurel$ there.
\item \textbf{Simulation.} Exact, by uniformisation, from a stationary initial state; 1{,}000 independent replicates per model (five batches of 200 with separate seeds), with the windows of each replicate nested.
\item \textbf{Prediction error $y$ (H1).} From each window, the passive prediction of the step response of $N$ to $\delta\ln\lambda$ is $\hat R(t)=\hat C_{NN}(0)-\hat C_{NN}(t)$, with $\hat C$ the sample autocovariance, i.e.\ the conjugate $B=N$ of \cref{prop:conjugate}. The error is
\[
y=\sup_{t\le3\tauint}\frac{|\hat R(t)-R_{\mathrm{exact}}(t)|}{\chi},
\]
where $R_{\mathrm{exact}}$ is the exact response and $\chi=\partial\avg{N}/\partial\ln\lambda$ the exact static response. In models B3, B2 and A the prediction is biased even with infinite data; the bias is reported as a floor.
\item \textbf{T-R (H2).} Dictionary: indicators of $N\in\{0\},\{1\},\dots,\{11\},\{\ge12\}$. Statistics: the 78 antisymmetric components $D_{ab}(\tau)$ of the two-time joint law of the binned $N$ at three lags $\tau\in\{\tauint/16,\tauint/8,\tauint/4\}$ (rounded to multiples of $\Delta$, at least $\Delta$), 234 in total. Components with fewer than 10 events in the window, or with a bootstrap standard error below $10^{-12}$, are excluded; without this guard, a few replicates with very short windows produced spurious statistics of order $10^{14}$ from floating-point standard errors of order $10^{-17}$. Each remaining component is standardised by its bootstrap standard error (non-overlapping blocks of $2\tauint$ plus the largest lag, 300 resamples), and the test statistic is the largest $|z|$. Its null distribution is obtained from the same bootstrap, centred at the observed values, and the test rejects at the 95\% quantile. The null needs neither a model nor a Markov assumption. \emph{Size} is measured on J, where $N$ is exactly reversible (\cref{thm:tandem}); \emph{power} on B3, B2 and A.
\item \textbf{Oracle calibration (diagnostic only).} To separate the calibration of the null from the strength of the signal, we also report the rejection rate with the threshold set at the empirical 95\% quantile of the statistic on J for the same $k$. This is not a usable test, since it needs a reversible reference system.
\item \textbf{Baseline.} A single cross-moment test, the antisymmetric part of $\E[u(N_0)v(N_\tau)]$ with $u$ and $v$ the standardised $N$ and $N^2$, at the same three lags, with the same block bootstrap and max-$|z|$ statistic. It is the simplest member of the cross-moment family of reversibility tests \citep{ramsey1996time}.
\item \textbf{Uncertainty.} Rejection rates are given with Wilson 95\% intervals \citep{wilson1927probable}; medians and 90th percentiles of $y$ with bootstrap 95\% intervals over replicates.
\end{itemize}

\subsection{Prediction (H1)}\label{sec:finite-pred}

\begin{table}[t]
\centering
\caption{Error $y$ of the passive prediction of the response of $N$ to $\delta\ln\lambda$ with $B=N$: median and 90th percentile over 1{,}000 replicates, with bootstrap 95\% intervals. The last row is the bias with infinite data.}
\label{tab:pred}
\small
\begin{tabular}{@{}rllrrr@{}}
\toprule
$k=\Tobs/\tauint$ & J: median & J: p90 & B3: median & B2: median & A: median\\
\midrule
10 & 0.50 [0.49, 0.52] & 1.00 [0.90, 1.13] & 0.52 & 0.55 & 1.13\\
30 & 0.33 [0.32, 0.35] & 0.67 [0.62, 0.78] & 0.38 & 0.41 & 1.64\\
100 & 0.20 [0.19, 0.21] & 0.43 [0.40, 0.45] & 0.23 & 0.27 & 1.80\\
300 & 0.12 [0.12, 0.13] & 0.25 [0.24, 0.27] & 0.15 & 0.20 & 1.88\\
1{,}000 & 0.07 [0.06, 0.07] & 0.15 [0.15, 0.16] & 0.12 & 0.18 & 1.91\\
3{,}000 & 0.04 [0.04, 0.04] & 0.09 [0.08, 0.09] & 0.11 & 0.17 & 1.91\\
10{,}000 & 0.02 [0.02, 0.02] & 0.05 [0.04, 0.05] & 0.11 & 0.17 & 1.90\\
\midrule
bias (infinite data) & 0 & 0 & 0.108 & 0.171 & 1.903\\
\bottomrule
\end{tabular}
\end{table}

When the conditions hold (model J), the error falls roughly as $k^{-1/2}$ (\cref{tab:pred}): the median is below 10\% from about $1{,}000\,\tauint$, and the 90th percentile from about $3{,}000\,\tauint$. When they fail, more data does not help: the error converges to the bias of the prediction, 11\% for B3, 17\% for B2 and 190\% for the autoscaler. The prediction is then consistently wrong, and only the tests can say so; this is why H1 and H2 have to be examined together.

\subsection{Detection with the reversibility test (H2)}\label{sec:finite-tr}

\begin{table}[t]
\centering
\caption{Rejection rate of T-R at nominal 5\% per window, with Wilson 95\% intervals (size on J, power on B3, B2, A); oracle-calibrated power and the rejection rate of the baseline cross-moment test (CM) in the last columns. 1{,}000 replicates per cell. Rows $k\le30$ are not valid tests (size far above 5\%) and are not reported as power.}
\label{tab:tr}
\resizebox{\textwidth}{!}{%
\begin{tabular}{@{}rlllllc@{}}
\toprule
$k$ & J: size & B3: power & B2: power & A: power & oracle (B3 / B2 / A) & CM (J / B3 / B2 / A)\\
\midrule
10 & 0.44 (invalid) & --- & --- & --- & --- & ---\\
30 & 0.16 (invalid) & --- & --- & --- & --- & ---\\
100 & 0.040 [0.030, 0.054] & 0.057 [0.044, 0.073] & 0.135 [0.115, 0.158] & 0.052 [0.040, 0.068] & 0.13 / 0.37 / 0.12 & 0 / 0.01 / 0.01 / 0\\
300 & 0.025 [0.017, 0.037] & 0.077 [0.062, 0.095] & 0.363 [0.334, 0.393] & 0.088 [0.072, 0.107] & 0.21 / 0.72 / 0.22 & 0 / 0.01 / 0 / 0\\
1{,}000 & 0.015 [0.009, 0.025] & 0.310 [0.282, 0.339] & 0.992 [0.984, 0.996] & 0.386 [0.356, 0.417] & 0.62 / 1.00 / 0.69 & 0 / 0 / 0 / 0\\
3{,}000 & 0.026 [0.018, 0.038] & 0.948 [0.932, 0.960] & 1.000 [0.996, 1.000] & 0.987 [0.978, 0.992] & 0.99 / 1.00 / 1.00 & 0 / 0.01 / 0.01 / 0\\
10{,}000 & 0.031 [0.022, 0.044] & 1.000 [0.996, 1.000] & 1.000 [0.996, 1.000] & 1.000 [0.996, 1.000] & 1.00 / 1.00 / 1.00 & 0 / 0.30 / 0.27 / 0\\
\bottomrule
\end{tabular}}
\end{table}

\begin{enumerate}[leftmargin=*]
\item \textbf{Validity.} With fewer than about 40 bootstrap blocks ($k\le30$) the null is miscalibrated (size 0.44 and 0.16; \cref{tab:tr}). From $k=100$ on, the size is between 1.5\% and 4\%, i.e.\ at or below the nominal 5\%. The minimum-count guard matters: in a first version of the study without it (200 replicates), the size was 0.5\% for most $k\ge300$; with it, the mean bootstrap threshold (3.5--3.6) is only 0.1--0.3 $z$ units above the empirical 95\% quantile of the statistic on J (3.3--3.4). This suggests that sparse components, whose bootstrap distributions are heavy-tailed, were the main source of conservativeness; longer blocks do not change the size much (\cref{sec:sensitivity}).
\item \textbf{Power.} B2 is detected with power 0.99 from $1{,}000\,\tauint$, B3 with power 0.95 at $3{,}000\,\tauint$, and the autoscaler with power 0.99 at $3{,}000\,\tauint$. Even the strong violation of model A needs about as long a window as the weak violation of model B3, because what T-R sees is the time asymmetry of $N$, which is moderate in both (\cref{tab:autoscaler}). With the oracle threshold the power at $1{,}000\,\tauint$ roughly doubles (0.62 for B3, 0.69 for A): the residual conservativeness of the bootstrap still costs up to about a factor of two in window length.
\item \textbf{Baseline.} The single cross-moment test has essentially no power below $10{,}000\,\tauint$ (0.30 and 0.27 for B3 and B2 there, none for A) and a size near zero. The indicator dictionary concentrates the signal where the asymmetry is, at specific pairs of levels of $N$, which a global moment dilutes.
\item \textbf{Wall-clock time.} Time is in units of the mean inter-arrival time of the tandem, so a window of $k\,\tauint$ scales inversely with the request rate. For B3, $3{,}000\,\tauint$ is about 51{,}000 inter-arrival times: about 8.5 minutes at 100 requests per second, and about 14 hours at one request per second. This is only indicative, since $\tauint$ of a real chain depends on its loads and structure; but the required windows are compatible with \cond{S} for busy services and may not be for lightly loaded ones.
\end{enumerate}

\subsection{The cost of the sufficiency test}\label{sec:tfcost}

We estimate the cost of a dynamic test of \cond{F} under an idealised design: $m$ load steps of size $\delta$ in $\ln\lambda$, each from a stationary state, are needed to resolve the difference between the measured and the predicted mean response of $N$ at a lag $t$, with per-step noise $\sqrt{2(\Var N-C_{NN}(t))}$ to leading order in $\delta$; $m$ is chosen for power 0.8 at a two-sided 5\% level and optimised over $t\le4\tauint$, and each step occupies its lag plus $3\tauint$ of re-equilibration. This is an order-of-magnitude estimate, not a bound: estimating the pre-step level from a longer window would reduce the noise by up to $\sqrt2$ (fewer steps), while the noise of the predicted response, ignored here, would increase it (more steps). Real control variation is also neither independent nor rectangular (\cref{sec:limitations}).

\begin{table}[t]
\centering
\caption{Number of load steps $m$ and telemetry needed by a dynamic T-F test, for the two forms of \cref{sec:testF}.}
\label{tab:tfcost}
\small
\begin{tabular}{@{}llrlrr@{}}
\toprule
Model & Conjugate & $\delta$ & Best lag & $m$ & Telemetry\\
\midrule
B3 & hypothesised, $B=N$ & 0.2 & $4.0\,\tauint$ & 2{,}356 & $\approx16{,}500\,\tauint$\\
B3 & estimated, $B^Y$ & 0.2 & $0.07\,\tauint$ & 82{,}309 & $\approx253{,}000\,\tauint$\\
B2 & hypothesised, $B=N$ & 0.2 & $4.0\,\tauint$ & 336 & $\approx2{,}400\,\tauint$\\
B2 & estimated, $B^Y$ & 0.2 & $0.02\,\tauint$ & 40{,}279 & $\approx122{,}000\,\tauint$\\
A & hypothesised, $B=N$ & 0.2 & $4.0\,\tauint$ & 34 & $\approx240\,\tauint$\\
A & estimated, $B^Y$ & 0.2 & $0.56\,\tauint$ & 304 & $\approx1{,}100\,\tauint$\\
\bottomrule
\end{tabular}
\end{table}

At $\delta=0.1$ every $m$ in \cref{tab:tfcost} is four times larger. The sufficiency test proper, with the estimated conjugate, is impractical for the weak violations of blocking, whose transient signature is small (\cref{prop:identifiability}(b)): it needs 35--120 times more steps than the joint test with the hypothesised conjugate. For the autoscaler it is affordable, about 300 steps of 20\% in load, and cheaper than T-R. The two tests are therefore complementary: T-R detects weak but persistent time asymmetries given long, finely sampled windows; T-F detects strong failures of sufficiency given routine control variation. With a hypothesised conjugate the joint test is cheaper still, but it tests the hypothesis together with \cond{F}.

\subsection{Sensitivity}\label{sec:sensitivity}

\begin{table}[t]
\centering
\caption{Rejection rate of T-R on the first 200 replicates, main configuration and five variants: sampling interval $\Delta=2$ and 8 (instead of 0.5; lags rounded to multiples of $\Delta$), 7 and 21 bins (instead of 13), blocks of $6\tauint$ (instead of $2\tauint$). Wilson 95\% intervals are about $\pm0.03$ for rates near 0.03 and $\pm0.07$ near 0.5.}
\label{tab:sens}
\small
\begin{tabular}{@{}lcc@{}}
\toprule
Variant & J size, $k=300$ / $1{,}000$ / $3{,}000$ & B3 power, $k=300$ / $1{,}000$ / $3{,}000$\\
\midrule
main & 0.030 / 0.005 / 0.010 & 0.085 / 0.330 / 0.925\\
$\Delta=2$ & 0.015 / 0.025 / 0.025 & 0.040 / 0.080 / 0.285\\
$\Delta=8$ & 0.035 / 0.020 / 0.030 & 0.035 / 0.015 / 0.050\\
7 bins & 0.030 / 0.015 / 0.030 & 0.140 / 0.555 / 0.995\\
21 bins & 0.025 / 0.005 / 0.010 & 0.065 / 0.265 / 0.845\\
blocks of $6\tauint$ & 0.060 / 0.025 / 0.020 & 0.140 / 0.405 / 0.920\\
\bottomrule
\end{tabular}
\end{table}

The size stays at or below about 5\% in every variant of \cref{tab:sens}. The power does not.
\begin{itemize}[leftmargin=*]
\item \textbf{The sampling interval is the critical design parameter.} Sampling every 2 time units ($\tauint/8.5$) instead of every 0.5 cuts the power at $3{,}000\,\tauint$ from 0.93 to 0.29, and sampling every 8 (about $\tauint/2$) removes it. The asymmetry of $N$ under blocking is concentrated at short lags, of the order of the service times, which coarse sampling cannot resolve, and coarser sampling also leaves fewer samples per window. Monitoring intervals of the order of the relaxation time, common in practice, would make this test blind to the violations studied here.
\item \textbf{A coarser dictionary helps.} Seven bins give power 0.56 at $1{,}000\,\tauint$ and 0.995 at $3{,}000$, against 0.33 and 0.93 with thirteen bins; twenty-one bins lose power. Fewer components mean fewer comparisons in the max-$|z|$ statistic and better-populated components.
\item \textbf{Longer blocks} change little (power 0.41 against 0.33 at $1{,}000\,\tauint$, size at most 6\%).
\end{itemize}

These numbers are for reference models, one family of dictionaries and one null construction. They give orders of magnitude for the windows, not universal constants, and they show which design choices matter.

\section{Time scales and levels}\label{sec:scales}

\subsection{The regime of a mechanism is set by its time scale}\label{sec:regimes}

``Top-level'' means that local mechanisms are integrated out. Which way they are integrated out depends on measured time scales, not on what the mechanism is called (\cref{tab:regimes}).

\begin{table}[t]
\centering
\caption{How a local mechanism enters the top level, by its time scale.}
\label{tab:regimes}
\small
\begin{tabularx}{\textwidth}{@{}lXX@{}}
\toprule
Regime & Treatment & Example\\
\midrule
$\tau_{\mathrm{mech}}\ll\taurel$ & \textbf{Averaged}: the mechanism enters through effective parameters & CPU throttling with quota periods much shorter than queue relaxation\\
$\tau_{\mathrm{mech}}\gg\taurel$ & \textbf{Quasi-static control}: TLOE holds window by window, and the equilibrium tracks the mechanism & A replica autoscaler acting every few minutes, with windows shorter than its period\\
$\tau_{\mathrm{mech}}\sim\taurel$ & \textbf{No separation}: the mechanism's state must enter $X$, possibly $\Phi$, and memory effects appear & An aggressively tuned autoscaler\\
\bottomrule
\end{tabularx}
\end{table}

Here $\tau_{\mathrm{mech}}$ is the time scale on which the mechanism's \emph{state} changes, which need not be its reaction delay: the hysteretic autoscaler of \cref{sec:autoscaler} keeps its replica count for times comparable to $\taurel$ however short its reaction delay, and is never averaged out.

The mapping between levels is that of nearly decomposable Markov systems \citep{simon1961aggregation,courtois1977decomposability}, developed for computer systems by Courtois, and of averaging for two-time-scale processes \citep{pavliotis2008multiscale,yin2013continuous}. With fast variables $z$ and slow variables $n$, effective slow rates are averages over the fast conditional law, $\mu_{\mathrm{eff}}(n)=\E_{\pi_{\mathrm{fast}}(z\mid n)}[\mu(n,z)]$ (aggregation), and given the slow state the fast state is distributed as $\pi_{\mathrm{fast}}(z\mid n)$ (disaggregation). Each level $l$ then has its own controls $\Theta_l$, top-level variables $\Phi_l$ and window, the levels communicate through effective parameters, and \cref{thm:resolution} applies level by level. Time-scale separation alone does not guarantee that the aggregated level inherits the properties of the detailed one: local detailed balance can fail after lumping despite a clear separation \citep{hartich2023violation}. This is one reason TLOE tests its conditions at each level rather than deriving them.

\subsection{Remark: sensitivities to fast controls \numerical}\label{sec:fastcontrols}

For a slow observable $A$ and a fast control $\theta_f$, when the slow level has a stationary law parametrised by an effective rate $\mu_{\mathrm{eff}}$, the chain rule through the averaged model reads
\begin{equation}\label{eq:chainrule}
\frac{\partial\avg{A}}{\partial\theta_f}\approx\frac{\partial\avg{A}}{\partial\mu_{\mathrm{eff}}}\,\frac{\partial\mu_{\mathrm{eff}}}{\partial\theta_f}=\Cov_{\mathrm{slow}}\big(A,\partial_{\mu_{\mathrm{eff}}}\ln\pi_{\mathrm{slow}}\big)\times\Cov_{\mathrm{fast}}\big(\mu(z),B_f(z)\big),
\end{equation}
where each factor is the static identity of \cref{prop:fdt} applied at its own level, and $B_f$ is the conjugate of $\theta_f$ in the fast stationary law. The content is not the chain rule itself but the operational reading: the sensitivity to a fast knob can be predicted from fluctuations measured separately at the two levels. The approximation is that of averaging. That sensitivities converge to those of the averaged model is known, at finite time \citep{gupta2014sensitivity} and in the stationary regime \citep{hashemi2016stochastic}, where an $O(\varepsilon)$ rate is stated to be open; we have not found the factorisation into two fluctuation--response covariances stated, and we do not prove a rate.

\paragraph{Check.} An M/M/1 queue ($\lambda=1$) has its service rate switched between 0.8 and 2.4 by a two-state process with rates $\gamma a$ (into the fast state) and $\gamma b$ (out of it), $a=b=1$, as for a server that is throttled part of the time. The control is $\theta_f=\ln a$. The rule predicts $\partial\avg{n}/\partial\theta_f=-\Var(n)/\mu_{\mathrm{eff}}\times(\mu_1-\mu_0)\,p(1-p)=-1.1111$ with $p=a/(a+b)$ and $\mu_{\mathrm{eff}}=1.6$. The exact sensitivity of the full chain is given in \cref{tab:chainrule}.

\begin{table}[t]
\centering
\caption{Exact sensitivity of the modulated M/M/1 queue to the fast control, and error of the rule \eqref{eq:chainrule}, which predicts $-1.1111$.}
\label{tab:chainrule}
\small
\begin{tabular}{@{}lrrrrr@{}}
\toprule
Switching speed $\gamma$ & 0.3 & 1 & 10 & 100 & 1000\\
\midrule
Exact $\partial\avg{n}/\partial\theta_f$ & $-2.263$ & $-1.513$ & $-1.157$ & $-1.1158$ & $-1.1116$\\
Relative error, $|\text{rule}-\text{exact}|/|\text{exact}|$ & 51\% & 27\% & 4.0\% & 0.42\% & 0.042\%\\
\bottomrule
\end{tabular}
\end{table}

Relative to the rule instead of the exact value, the errors are 104\%, 36\%, 4.1\%, 0.42\% and 0.042\%. On this example the error decays as $1/\gamma$, i.e.\ linearly in the ratio of time scales. Without separation ($\gamma\lesssim1$) the rule fails, as the third regime of \cref{tab:regimes} says it should.

\section{Related work}\label{sec:related}

\paragraph{Equilibrium in computing-continuum systems.} TLOE continues a line in which equilibrium was introduced as a notion for managing computing-continuum applications: informally, as an operation mode tied to an infrastructure configuration and a way to move away from thresholds \citep{casamayorpujol2021towards,dustdar2023distributed}; as the target state of agents acting through active inference \citep{sedlak2024equilibrium}; and formally, as the fulfilment of all SLOs \citep{casamayorpujol2025distributed}.
In all three, equilibrium is a statement about SLOs: a system is in equilibrium when it meets its objectives. TLOE is a statement about the operational state: a system is in equilibrium when the stochastic behaviour of its state satisfies conditions that can be tested on telemetry, whether or not its objectives are met.
Closer to TLOE is the intuition of \citet{rouska2025equilibrium} that equilibrium is predictive regulation under perturbation, measured by a KL divergence. TLOE differs from it in three ways: it is a property of the system's law rather than of the pair (system, internal model), so agreement between a model and the system is not equilibrium (\cref{sec:kl}); it has a time-reversal condition, without which a circulating current goes undetected; and the FDT is a theorem whose failure can be diagnosed, rather than an analogy. Outside this line, stability of elastic resource management is a property of the control loop \citep{loridobotran2014review}, and steady state is detected empirically \citep{franklin2008stationarity}; neither relates the notion to the response of the system.

\paragraph{Response theory in non-equilibrium steady states.} The identity of \cref{prop:fdt} holds in any Markov steady state \citep{prost2009generalized,seifert2010fluctuation}, and under detailed balance goes back to \citet{agarwal1972fluctuation}. The class of perturbation fixes the conjugate: perturbations of state energies give an exact, equilibrium-form static FDT with a local conjugate in any NESS, and verifying such an FDT does not establish equilibrium \citep{owen2020universal}; for diffusions, see \citet{gao2022thermodynamic}. Responses are bounded by unperturbed fluctuations \citep{dechant2020fluctuation}, and can be computed from unperturbed trajectories through correlations of local transitions \citep{zheng2025unified}, including for non-Markovian dynamics, where \emph{response-sufficient} coarse variables are defined \citep{zheng2026memory}; the latter is the closest notion to the dynamic content of \cond{F}. The response to a parameter change cannot be identified from data at a single condition, even with fully observed dynamics \citep{thommen2026limits}, which is why TLOE requires the controls to be declared.

\paragraph{Marginal observers: the foundation.} The closest prior theory is that of marginal observers \citep{polettini2017effective,polettini2019effective}. An observer who measures a subset of the currents of a system has effective affinities for which a marginal fluctuation relation holds; tuning parameters until the observed currents stall makes the hidden time-reversal generator coincide with the forward one, the analogue of detailed balance; and bringing some currents to stalling perturbs others. Equilibrium-form response relations relative to an observer also hold when the perturbation touches only the observed transitions \citep{altaner2016fluctuation} or around stalling \citep{shiraishi2022time}. Three points of contact need to be stated precisely.
\begin{itemize}[leftmargin=*]
\item \emph{What the observer sees and controls.} In \citet{polettini2019effective} the observer measures a chosen subset of currents (edge currents or phenomenological currents, i.e.\ linear combinations of edge currents), and their Assumption B0 requires the thermodynamic parameters to act only on the edges that support the observed currents \citep[Sec.~II.B]{polettini2019effective}. This ties parameters to observables, as \cref{prop:conjugate} does, but in the opposite direction: B0 restricts the parameters to fit the chosen observables, while \cond{F} fixes the declared controls and derives the admissible observables from sufficiency of the stationary law. We have not found a sufficiency condition in their framework.
\item \emph{Marginal consistency.} For phenomenological currents, their theory requires \emph{marginal consistency} (their Prop.~29, Eq.~258): the effective affinities of the edges that carry the observed currents must be linear combinations of the phenomenological effective affinities. Under it the hidden time-reversed generator is a Markov generator (Prop.~30), and a phenomenological stalling state is also internally stalling (Prop.~31); conversely, if the marginal entropy production vanishes at a phenomenological stalling state, marginal consistency holds (Prop.~32). Marginal consistency is strictly stronger than complete thermodynamic consistency, and the gap is the open question the authors single out (Sec.~VI.E). It is a condition on the affinities of currents, analogous to local detailed balance; \cond{R} is a condition on the time-reversal symmetry of a state aggregate. We know of no implication between them in either direction.
\item \emph{Stalling versus \cond{R}.} Their equilibrium-like symmetries (the hidden time-reversal generator coinciding with the forward one, the symmetrised fluctuation--dissipation relation) hold at stalling, where the observed currents vanish; in a service chain that means stopping the service. \cond{R} concerns a state aggregate, which in the tandem is reversible at full throughput (\cref{thm:tandem}).
\end{itemize}
By its authors' account (Sec.~II.C), it is not a kinetic theory: it does not coarse-grain the hidden dynamics into an effective marginal dynamics representative of finite-time behaviour (rate of convergence to the steady state, first exit times), it does not involve time-scale separation between marginal and hidden degrees of freedom, and it leaves open whether a phenomenological procedure can determine the hidden time-reversed dynamics. TLOE supplies an operational version of these pieces: \cond{S} makes time-scale separation explicit and partly testable, the relaxation spectrum and implied time scales give the kinetics, and \cref{sec:tests} gives a data-driven procedure.

\paragraph{Coarse-graining and lumpability.} Fast thermalisation inside macrostates recovers the microscopic structure at the coarse level \citep{esposito2012stochastic}; decimating a chain without detailed balance can yield a reversible one \citep{puglisi2010entropy}; exact coarse-graining preserves entropy production \citep{teza2020exact}; local detailed balance can fail after lumping despite time-scale separation \citep{hartich2023violation}; time reversal and coarse-graining do not commute, so coarse observation can mask or distort broken time-reversal symmetry \citep{dieball2025perspective}; projected coarse models can satisfy an FDT by construction and still predict the wrong response \citep{jung2022nonmarkovian}; and hidden entropy production can only be bounded from coarse data \citep{martinez2019inferring}. Sufficient statistics of coarse trajectories have been used in the opposite direction to TLOE, to \emph{detect} hidden non-equilibrium \citep{wu2025parameter}. In the theory of Markov chains, lumpability asks when a function of a chain is itself Markov \citep{kemeny1960finite}, and reversibility relative to a lumping has been defined for strictly lumpable chains \citep{marin2017relations}, without parameters or controls. TLOE needs neither: the total WIP of the tandem is not lumpable (\cref{sec:consequences}), yet it is reversible and sufficient. Lumped detailed balance \eqref{eq:lumpedDB}, used in \cref{sec:autoscaler}, is the necessary condition for reversibility of a coarse-grained process that remains when lumpability fails. Apparent equilibrium at the top level can therefore be genuine or an artefact of the choice of $Y$; \cond{F}, which ties $Y$ to the controls, the transient test of \cref{sec:testF}, and the Markovianity check of \cref{sec:testR} are the protections TLOE adds.

\paragraph{Queueing networks.} Product-form laws are exponential families, with the score used for policy gradients \citep{comte2025score}; product form coincides with complex balance in deficiency-zero reaction networks \citep{anderson2010product}; open queueing networks map onto zero-range processes and have been studied as driven non-equilibrium systems, including current large deviations \citep{chernyak2010nonequilibrium}; and in the open zero-range process an equal shift of both entrance rates gives an equilibrium-form response with the total particle number as conjugate \citep{maes2014linear}, the closest physical analogue of \cref{prop:conjugate}(a). Sensitivities can be estimated from a single sample path by likelihood-ratio methods and perturbation analysis \citep{glynn1990likelihood,ho1991perturbation,cao1997perturbation}, which need event-level paths, and configuration changes can be predicted from deployment traces that already cover the scenario \citep{tariq2008answering}. TLOE differs in working from coarse top-level telemetry, in not requiring the scenario to be covered by past data, and in carrying a test of when its prediction is valid. \Cref{lem:tandem,lem:block} rest on the interchangeability of $\cdot$/M/1 queues \citep{weber1979interchangeability,tsoucas1987interchangeability} and on the time reversal of open networks \citep{kelly1979reversibility}. In that literature ``reversibility'' often means reversing the order of the stations \citep[e.g.][]{chao1992reversibility}; here it always means time-reversibility.

\paragraph{Identification and monitoring.} The fluctuation--response identity is a non-parametric estimator of a step response from passive operating data, and the limits of \cref{prop:identifiability} have a counterpart in system identification, where a reduced model can match the static gain and miss the transient \citep{ljung1999system}. Testing \cond{S} and running the tests continuously draws on stationarity and change-point testing \citep{kwiatkowski1992testing,page1954continuous} and on sequential testing \citep{wald1945sequential}.

\paragraph{Other equilibria.} Network utility maximisation \citep{kelly1997charging,kelly1998rate}, Wardrop equilibria and mean-field-game equilibria are equilibria of optimisation or of strategic interaction. They answer different questions from TLOE, which is a property of the stochastic dynamics of a running system relative to its controls.

\paragraph{Practice.} \Cref{tab:practice} places TLOE among the methods operators use.

\begin{table}[t]
\centering
\caption{Methods for assessing or predicting the behaviour of a running distributed system.}
\label{tab:practice}
\footnotesize
\begin{tabularx}{\textwidth}{@{}>{\raggedright\arraybackslash}p{3.4cm}>{\raggedright\arraybackslash}X>{\raggedright\arraybackslash}p{2.6cm}>{\raggedright\arraybackslash}X@{}}
\toprule
Family & Detects / predicts & Interventions & Main limitation\\
\midrule
SLO alerting and thresholds \citep{thurgood2018alerting,loridobotran2014review} & That a target is being violated & None & Reactive; thresholds hard to tune and can cause oscillation\\
AIOps and root-cause analysis \citep{notaro2021survey,soldani2022anomaly} & Which component caused a failure & None & Diagnoses failures, does not predict responses\\
Canary and chaos experiments \citep{warner2018canarying,basiri2019automating} & Whether an applied change caused a regression & Yes & Cost and exposure; answers only the question asked\\
Model-based prediction (queueing, architecture-level, learned) \citep{brosig2015quantitative,guemespalau2026simulation} & KPIs under what-if changes & None on production & Valid only under the model's assumptions or near training data\\
System identification from operating data \citep{ljung1999system} & A dynamic model of the response & Excitation or routine variation & Model structure chosen in advance; validity under feedback and drift\\
Passive sensitivity estimation \citep{cao1997perturbation,tariq2008answering} & Effect of a change without applying it & None & Event-level paths, or coverage of the scenario\\
Early-warning signals \citep{dabrowski2018evaluating} & Approach to collapse & None & Did not outperform thresholds in realistic simulations\\
\textbf{TLOE} & Whether the response to a control change is predictable, and the response & None for T-S, T-R and the prediction; routine control variation for T-F & Exact only conditionally on \cond{S}, \cond{F}, \cond{R}; long windows (\cref{sec:finite}); evaluated on reference models only\\
\bottomrule
\end{tabularx}
\end{table}

\section{Limitations and open problems}\label{sec:limitations}

\begin{enumerate}[leftmargin=*]
\item \textbf{No real telemetry.} Every number in this paper comes from reference models: two- to five-stage tandems, two blocking variants and one autoscaler. Whether real edge--cloud services satisfy \cond{S} at all, i.e.\ whether $\taulocal\ll\taurel\ll\Tobs\ll\taudrift$ holds for their workloads, and whether the predictions survive real telemetry, are the questions this paper does not answer; the window lengths of \cref{sec:finite} suggest that busy services are the natural first target.
\item \textbf{The sufficiency test is specified, not run.} \Cref{sec:tfcost} estimates its cost under an idealised design of independent load steps from stationarity; real control variation is neither independent nor rectangular, and is confounded with the natural variation of the load. A practical T-F, and its size and power, remain to be established.
\item \textbf{Beyond exponential service.} With phase-type service the operational state stays Markov once the phase of each job in service is added to it; with general service times it is a generalised semi-Markov process. Whether \cond{F} survives at the total WIP then depends on the service discipline. For symmetric disciplines such as processor sharing, the stationary law of the extended state is the product-form law of the counts times a phase factor that does not depend on the arrival rates \citep[Sec.~3.3, Theorems~3.8 and~3.10]{kelly1979reversibility}, so the conjugate of the load is still a function of the counts: \cond{F} holds at the total WIP, and the prediction stays exact over time, not only in the static limit. A single queue with Erlang-2 service at load 0.7 confirms this: the fraction of the conjugate's variance not explained by $n$ is below $10^{-8}$, and the predicted step response matches the exact one to $6\times10^{-11}$ of the static response. Under FCFS the phase carries part of the conjugate: $n$ leaves 0.5\% of its variance unexplained, and the static response predicted with $B=n$ is 20\% too low (4.81 against 6.01) \numerical. Interchangeability fails for non-exponential FCFS stations (\cref{sec:tandem}, Erlang check), so \cref{thm:tandem,cor:stage} do not extend as stated; whether total WIP is time-reversible in chains of processor-sharing stations is open.
\item \textbf{More counter-models.} Blocking (after service only) and one hysteretic autoscaler have been computed. Retries modelled as a Hawkes process, shared resources, fork--join stages and closed populations are the next ones; the claim that retry storms appear as violations of \cond{S} remains a hypothesis. For the autoscaler, the non-existence of an admissible aggregate is established numerically at five reaction times (\cref{sec:autoscaler}); a proof for every $\alpha>0$ is open.
\item \textbf{Networks beyond tandems.} Whether total WIP is time-reversible in feed-forward Jackson networks with branching and merging is open; \cref{lem:tandem,lem:block} cover chains only. Whether the condition of \cref{cor:stage}(b) is also necessary in general is open (it is in every case computed).
\item \textbf{Relation of \cond{R} to marginal consistency} in the theory of marginal observers (\cref{sec:related}).
\item \textbf{A finite measure of irreversibility.} The entropy production of a directed workflow at stage resolution is infinite; a finite measure, such as the relative-entropy rate of $Y$, and its quantitative link to the error of the prediction, as \citet{harada2005equality} give for Langevin systems, are open. Current large deviations of queueing networks \citep{chernyak2010nonequilibrium} and the thermodynamics of complex-balanced networks \citep{rao2016nonequilibrium} are natural starting points.
\item \textbf{Statistics.} The block-bootstrap null of \cref{sec:finite} is conservative; better-calibrated nulls, adaptive dictionaries, a baseline beyond the single cross-moment test, sequential procedures for continuous monitoring, and tests that keep their power when the monitoring interval is comparable to $\tauint$ (\cref{sec:sensitivity} shows that the present test loses it) are open. The window lengths of \cref{sec:finite} are for reference models and one family of tests.
\item \textbf{Time-dependent environments.} Daily cycles, cron jobs and deadlines break \cond{S}; a cyclostationary version of TLOE is a candidate extension.
\item \textbf{Closing the operational loop.} The paper predicts the response of top-level observables to small control changes and says when the prediction can be trusted. How an operator should combine that prediction with SLOs, risk and the frequency of adaptation (predict, decide, act, re-test) is not addressed.
\end{enumerate}

\section{Conclusion}\label{sec:conclusion}

We have proposed a definition of equilibrium for a running distributed system that is relative to the controls its operator declares and testable, in part, on the telemetry the system already produces. Its three conditions, stationarity with time-scale separation, sufficiency of a top-level aggregate for the controls, and time-reversibility of that aggregate, reduce to two: stationarity, and time-reversibility of the minimal sufficient statistic of the controls; declaring fewer controls can only help. Under the first two conditions, the response to a small control change follows from passive telemetry with a conjugate fixed by the type of control. A chain of exponential servers is in this equilibrium for load control at every stable throughput, although every job flows one way; adding a capacity control at the first or last stage takes it out, while a control at the middle stage of a balanced chain does not; blocking weakens it; and a hysteretic autoscaler destroys it in every case computed. The tests that decide which case applies need long windows: the reversibility test needs a few thousand integrated autocorrelation times, finely sampled, to detect the violations studied here, and a test of sufficiency is prohibitively expensive for weak violations but affordable for strong ones. Equilibrium in this sense is not health: it is the property that makes the response of a system to adaptation predictable, and that tells the operator when a prediction cannot be trusted.

\paragraph{Code and data availability.} Every number in this paper is produced by Python scripts, listed with their outputs in \cref{app:numerics}; the stored outputs, including the per-replicate statistics of \cref{sec:finite}, are part of the repository, available at \url{https://github.com/vikcas/TLOE-theoretical_analysis} and archived on Zenodo (version~1.0.0, \href{https://doi.org/10.5281/zenodo.24200155}{doi:10.5281/zenodo.24200155}), under the MIT licence.

\paragraph{Use of AI tools.} The authors used Claude (Anthropic) to assist with drafting and editing the text, checking derivations, writing the verification and simulation scripts, and checking references. The authors reviewed and verified all content and take full responsibility for it.

\bibliographystyle{plainnat}
\bibliography{tloe_references}

@article{agarwal1972fluctuation,
  author = {Agarwal, G. S.}, title = {Fluctuation-dissipation theorems for systems in non-thermal equilibrium and applications},
  journal = {Zeitschrift f{\"u}r Physik}, volume = {252}, number = {1}, pages = {25--38}, year = {1972}, doi = {10.1007/BF01391621}
}

@article{altaner2016fluctuation,
  author = {Altaner, Bernhard and Polettini, Matteo and Esposito, Massimiliano}, title = {Fluctuation-Dissipation Relations Far from Equilibrium},
  journal = {Physical Review Letters}, volume = {117}, pages = {180601}, year = {2016}, eprint = {1604.08832}, archivePrefix = {arXiv}, number = {18}, doi = {10.1103/PhysRevLett.117.180601}
}

@article{anderson2010product,
  author = {Anderson, David F. and Craciun, Gheorghe and Kurtz, Thomas G.}, title = {Product-form stationary distributions for deficiency zero chemical reaction networks},
  journal = {Bulletin of Mathematical Biology}, volume = {72}, pages = {1947--1970}, year = {2010}, eprint = {0803.3042}, archivePrefix = {arXiv}, number = {8}, doi = {10.1007/s11538-010-9517-4}
}

@article{antoch2026experimental,
  author = {Antoch, Jarom{\'\i}r and Binder, Walter and Bulej, Lubom{\'\i}r and Farquet, Fran{\c{c}}ois and Hork{\'y}, Vojt{\v{e}}ch and Prokopec, Aleksandar and Ros{\`a}, Andrea and T{\r{u}}ma, Petr}, title = {Experimental Evaluation Methodology for the Era of No Steady Performance},
  journal = {Proceedings of the ACM on Programming Languages}, volume = {10}, number = {OOPSLA2}, pages = {316:1--316:28}, month = oct, year = {2026}, doi = {10.1145/3839448}
}

@misc{basiri2019automating,
  author = {Basiri, Ali and Hochstein, Lorin and Jones, Nora and Tucker, Haley}, title = {Automating chaos experiments in production},
  year = {2019}, eprint = {1905.04648}, howpublished = {arXiv:1905.04648}, archivePrefix = {arXiv}, doi = {10.48550/arXiv.1905.04648}
}

@article{baskett1975open,
  author = {Baskett, Forest and Chandy, K. Mani and Muntz, Richard R. and Palacios, Fernando G.}, title = {Open, Closed, and Mixed Networks of Queues with Different Classes of Customers},
  journal = {Journal of the ACM}, volume = {22}, number = {2}, pages = {248--260}, year = {1975}, doi = {10.1145/321879.321887}
}

@inproceedings{boyen1998tractable,
  author = {Boyen, Xavier and Koller, Daphne}, title = {Tractable inference for complex stochastic processes},
  booktitle = {Proceedings of the Fourteenth Conference on Uncertainty in Artificial Intelligence (UAI)}, pages = {33--42}, publisher = {Morgan Kaufmann}, address = {San Francisco}, year = {1998}
}

@article{brosig2015quantitative,
  author = {Brosig, Fabian and Meier, Philipp and Becker, Steffen and Koziolek, Anne and Koziolek, Heiko and Kounev, Samuel}, title = {Quantitative Evaluation of Model-Driven Performance Analysis and Simulation of Component-Based Architectures},
  journal = {IEEE Transactions on Software Engineering}, volume = {41}, number = {2}, pages = {157--175}, year = {2015}, doi = {10.1109/TSE.2014.2362755}
}

@article{cao1997perturbation,
  author = {Cao, Xi-Ren and Chen, Han-Fu}, title = {Perturbation realization, potentials, and sensitivity analysis of Markov processes},
  journal = {IEEE Transactions on Automatic Control}, volume = {42}, number = {10}, pages = {1382--1393}, year = {1997}, doi = {10.1109/9.633827}
}

@book{casella2002statistical,
  author = {Casella, George and Berger, Roger L.}, title = {Statistical Inference}, edition = {2nd}, publisher = {Duxbury}, address = {Pacific Grove, CA}, year = {2002}
}

@inproceedings{casamayorpujol2021towards,
  author = {Casamayor Pujol, V{\'\i}ctor and Raith, Philipp and Dustdar, Schahram}, title = {Towards a new paradigm for managing computing continuum applications},
  booktitle = {2021 IEEE Third International Conference on Cognitive Machine Intelligence (CogMI)}, year = {2021}, doi = {10.1109/cogmi52975.2021.00032}
}

@misc{casamayorpujol2025distributed,
  author = {Casamayor Pujol, V{\'\i}ctor and Sedlak, Boris and Salvatori, Tommaso and Friston, Karl and Dustdar, Schahram}, title = {Distributed Intelligence in the Computing Continuum with Active Inference},
  year = {2025}, eprint = {2505.24618}, howpublished = {arXiv:2505.24618}, archivePrefix = {arXiv}, doi = {10.48550/arXiv.2505.24618}
}

@article{chao1992reversibility,
  author = {Chao, Xiuli and Pinedo, Michael}, title = {On reversibility of tandem queues with blocking},
  journal = {Naval Research Logistics}, volume = {39}, number = {7}, pages = {957--974}, year = {1992}, doi = {10.1002/1520-6750(199212)39:7<957::AID-NAV3220390707>3.0.CO;2-3}
}

@article{chernyak2010nonequilibrium,
  author = {Chernyak, Vladimir Y. and Chertkov, Michael and Goldberg, David A. and Turitsyn, Konstantin}, title = {Non-equilibrium statistical physics of currents in queuing networks},
  journal = {Journal of Statistical Physics}, volume = {140}, pages = {819--845}, year = {2010}, doi = {10.1007/s10955-010-0018-5}, eprint = {1001.5454}, archivePrefix = {arXiv}, number = {5}
}

@article{comte2025score,
  author = {Comte, C{\'e}line and Jonckheere, Matthieu and Sanders, Jaron and Senen-Cerda, Albert}, title = {Score-Aware Policy-Gradient and Performance Guarantees using Local Lyapunov Stability},
  journal = {Journal of Machine Learning Research}, volume = {26}, number = {132}, pages = {1--74}, year = {2025}
}

@book{courtois1977decomposability,
  author = {Courtois, Pierre-Jacques}, title = {Decomposability: Queueing and Computer System Applications}, publisher = {Academic Press}, address = {New York}, year = {1977}
}

@inproceedings{dabrowski2018evaluating,
  author = {Dabrowski, Christopher and Mills, Kevin}, title = {Evaluating predictors of congestion collapse in communication networks},
  booktitle = {NOMS 2018 - 2018 IEEE/IFIP Network Operations and Management Symposium}, pages = {1--9}, year = {2018}, doi = {10.1109/NOMS.2018.8406225}
}

@article{dechant2020fluctuation,
  author = {Dechant, Andreas and Sasa, Shin-ichi}, title = {Fluctuation--response inequality out of equilibrium},
  journal = {Proceedings of the National Academy of Sciences}, volume = {117}, number = {12}, pages = {6430--6436}, year = {2020}, doi = {10.1073/pnas.1918386117}
}

@article{dieball2025perspective,
  author = {Dieball, Cai and Godec, Alja}, title = {Perspective: Time irreversibility in systems observed at coarse resolution},
  journal = {The Journal of Chemical Physics}, volume = {162}, number = {9}, pages = {090901}, year = {2025}, doi = {10.1063/5.0251089}
}

@article{dustdar2023distributed,
  author = {Dustdar, Schahram and Casamayor Pujol, V{\'\i}ctor and Donta, Praveen Kumar}, title = {On Distributed Computing Continuum Systems},
  journal = {IEEE Transactions on Knowledge and Data Engineering}, volume = {35}, number = {4}, pages = {4092--4105}, year = {2023}, doi = {10.1109/TKDE.2022.3142856}
}

@article{edwards1979large,
  author = {Edwards, David G.}, title = {Large Sample Tests for Stationarity and Reversibility in Finite {M}arkov Chains},
  journal = {Scandinavian Journal of Statistics}, volume = {7}, number = {4}, pages = {203--206}, year = {1980}, url = {http://www.jstor.org/stable/4615801}
}

@article{esposito2012stochastic,
  author = {Esposito, Massimiliano}, title = {Stochastic thermodynamics under coarse graining},
  journal = {Physical Review E}, volume = {85}, number = {4}, pages = {041125}, year = {2012}, doi = {10.1103/PhysRevE.85.041125}
}

@inproceedings{franklin2008stationarity,
  author = {Franklin, W. Wesley and White, K. Preston}, title = {Stationarity tests and {MSER-5}: Exploring the intuition behind mean-squared-error-reduction in detecting and correcting initialization bias},
  booktitle = {2008 Winter Simulation Conference}, pages = {541--546}, year = {2008}, doi = {10.1109/WSC.2008.4736111}
}

@article{gao2022thermodynamic,
  author = {Gao, Qi and Chun, Hyun-Myung and Horowitz, Jordan M.}, title = {Thermodynamic constraints on the nonequilibrium response of one-dimensional diffusions},
  journal = {Physical Review E}, volume = {105}, pages = {L012102}, year = {2022}, doi = {10.1103/PhysRevE.105.L012102}, eprint = {2112.05796}, archivePrefix = {arXiv}
}

@article{glynn1990likelihood,
  author = {Glynn, Peter W.}, title = {Likelihood ratio gradient estimation for stochastic systems},
  journal = {Communications of the ACM}, volume = {33}, number = {10}, pages = {75--84}, year = {1990}, doi = {10.1145/84537.84552}
}

@incollection{goto2026modelfree,
  author = {Goto, Yuichi and Hallin, Marc}, title = {A model-free test of the time-reversibility of climate change processes},
  booktitle = {Asymptotic and Methodological Statistics: Festschrift in Honor of Marie Hu{\v{s}}kov{\'a}}, editor = {Hlubinka, Daniel and Hudecov{\'a}, {\v{S}}{\'a}rka and Maciak, Mat{\'u}{\v{s}} and Pe{\v{s}}ta, Michal},
  publisher = {Springer}, pages = {45--58}, year = {2026}, doi = {10.1007/978-3-032-07178-1_3}, eprint = {2411.11248}, archivePrefix = {arXiv}
}

@article{guemespalau2026simulation,
  author = {G{\"u}emes-Palau, Carlos and Ferriol-Galm{\'e}s, Miquel and Paillisse-Vilanova, Jordi and Barlet-Ros, Pere and Cabellos-Aparicio, Albert}, title = {From simulation to deep learning: Survey on network performance modeling approaches},
  journal = {Computer Networks}, volume = {282}, year = {2026}, doi = {10.1016/j.comnet.2026.112253}
}

@article{gupta2014sensitivity,
  author = {Gupta, Ankit and Khammash, Mustafa}, title = {Sensitivity analysis for stochastic chemical reaction networks with multiple time-scales},
  journal = {Electronic Journal of Probability}, volume = {19}, pages = {1--53}, year = {2014}, doi = {10.1214/EJP.v19-3246}
}

@article{harada2005equality,
  author = {Harada, Takahiro and Sasa, Shin-ichi}, title = {Equality connecting energy dissipation with a violation of the fluctuation-response relation},
  journal = {Physical Review Letters}, volume = {95}, pages = {130602}, year = {2005}, number = {13}, doi = {10.1103/PhysRevLett.95.130602}
}

@article{hartich2023violation,
  author = {Hartich, David and Godec, Alja}, title = {Violation of local detailed balance upon lumping despite a clear timescale separation},
  journal = {Physical Review Research}, volume = {5}, number = {3}, pages = {L032017}, year = {2023}, doi = {10.1103/PhysRevResearch.5.L032017}
}

@article{hashemi2016stochastic,
  author = {Hashemi, Araz and N{\'u}{\~n}ez, Marcel and Plech{\'a}{\v{c}}, Petr and Vlachos, Dionisios G.}, title = {Stochastic averaging and sensitivity analysis for two scale reaction networks},
  journal = {The Journal of Chemical Physics}, volume = {144}, pages = {074104}, year = {2016}, eprint = {1509.03802}, archivePrefix = {arXiv}, number = {7}, doi = {10.1063/1.4942008}
}

@article{hawkes1971spectra,
  author = {Hawkes, Alan G.}, title = {Spectra of some self-exciting and mutually exciting point processes},
  journal = {Biometrika}, volume = {58}, number = {1}, pages = {83--90}, year = {1971}, doi = {10.1093/biomet/58.1.83}
}

@book{ho1991perturbation,
  author = {Ho, Yu-Chi and Cao, Xi-Ren}, title = {Perturbation Analysis of Discrete Event Dynamic Systems}, publisher = {Kluwer Academic Publishers}, address = {Boston}, year = {1991},
  series = {The Springer International Series in Engineering and Computer Science}, volume = {145}, doi = {10.1007/978-1-4615-4024-3}
}

@article{jackson1963jobshop,
  author = {Jackson, James R.}, title = {Jobshop-Like Queueing Systems}, journal = {Management Science}, volume = {10}, number = {1}, pages = {131--142}, year = {1963}, doi = {10.1287/mnsc.10.1.131}
}

@article{jung2022nonmarkovian,
  author = {Jung, Gerhard}, title = {Non-{M}arkovian systems out of equilibrium: Exact results for two routes of coarse graining},
  journal = {Journal of Physics: Condensed Matter}, volume = {34}, number = {20}, pages = {204004}, year = {2022}, doi = {10.1088/1361-648X/ac56a7}
}

@book{kelly1979reversibility,
  author = {Kelly, Frank P.}, title = {Reversibility and Stochastic Networks}, publisher = {Wiley}, address = {Chichester}, year = {1979}
}

@article{kelly1997charging,
  author = {Kelly, Frank P.}, title = {Charging and rate control for elastic traffic}, journal = {European Transactions on Telecommunications}, volume = {8}, pages = {33--37}, year = {1997}, number = {1}, doi = {10.1002/ett.4460080106}
}

@article{kelly1998rate,
  author = {Kelly, Frank P. and Maulloo, Aman K. and Tan, David K. H.}, title = {Rate control for communication networks: shadow prices, proportional fairness and stability},
  journal = {Journal of the Operational Research Society}, volume = {49}, pages = {237--252}, year = {1998}, number = {3}, doi = {10.1038/sj.jors.2600523}
}

@book{kemeny1960finite,
  author = {Kemeny, John G. and Snell, J. Laurie}, title = {Finite {M}arkov Chains}, publisher = {Van Nostrand}, address = {Princeton}, year = {1960}
}

@article{kwiatkowski1992testing,
  author = {Kwiatkowski, Denis and Phillips, Peter C. B. and Schmidt, Peter and Shin, Yongcheol}, title = {Testing the null hypothesis of stationarity against the alternative of a unit root},
  journal = {Journal of Econometrics}, volume = {54}, pages = {159--178}, year = {1992}, number = {1--3}, doi = {10.1016/0304-4076(92)90104-Y}
}

@article{lehmann1950completeness,
  author = {Lehmann, Erich L. and Scheff{\'e}, Henry}, title = {Completeness, similar regions, and unbiased estimation: Part {I}},
  journal = {Sankhy{\=a}: The Indian Journal of Statistics}, volume = {10}, number = {4}, pages = {305--340}, year = {1950}
}

@book{ljung1999system,
  author = {Ljung, Lennart}, title = {System Identification: Theory for the User}, edition = {2nd}, publisher = {Prentice Hall PTR}, address = {Upper Saddle River, NJ}, year = {1999},
  series = {Prentice Hall Information and System Sciences Series}, isbn = {9780136566953}
}

@article{loridobotran2014review,
  author = {Lorido-Botran, Tania and Miguel-Alonso, Jose and Lozano, Jose A.}, title = {A Review of Auto-scaling Techniques for Elastic Applications in Cloud Environments},
  journal = {Journal of Grid Computing}, volume = {12}, number = {4}, pages = {559--592}, year = {2014}, doi = {10.1007/s10723-014-9314-7}
}

@article{maes2014linear,
  author = {Maes, Christian and Salazar, Alberto}, title = {Linear response in the nonequilibrium zero range process},
  journal = {Chaos, Solitons \& Fractals}, volume = {64}, pages = {78--87}, year = {2014}, doi = {10.1016/j.chaos.2013.09.004}
}

@article{marin2017relations,
  author = {Marin, Andrea and Rossi, Sabina}, title = {On the relations between {M}arkov chain lumpability and reversibility},
  journal = {Acta Informatica}, volume = {54}, number = {5}, pages = {447--485}, year = {2017}, doi = {10.1007/s00236-016-0266-1}
}

@article{martinez2019inferring,
  author = {Mart{\'\i}nez, Ignacio A. and Bisker, Gili and Horowitz, Jordan M. and Parrondo, Juan M. R.}, title = {Inferring broken detailed balance in the absence of observable currents},
  journal = {Nature Communications}, volume = {10}, pages = {3542}, year = {2019}, doi = {10.1038/s41467-019-11051-w}
}

@article{morse1955stochastic,
  author = {Morse, Philip M.}, title = {Stochastic properties of waiting lines}, journal = {Journal of the Operations Research Society of America}, volume = {3}, number = {3}, pages = {255--261}, year = {1955}, doi = {10.1287/opre.3.3.255}
}

@article{murashita2014nonequilibrium,
  author = {Murashita, Yuto and Funo, Ken and Ueda, Masahito}, title = {Nonequilibrium equalities in absolutely irreversible processes},
  journal = {Physical Review E}, volume = {90}, pages = {042110}, year = {2014}, number = {4}, doi = {10.1103/PhysRevE.90.042110}
}

@article{notaro2021survey,
  author = {Notaro, Paolo and Cardoso, Jorge and Gerndt, Michael}, title = {A Survey of {AIOps} Methods for Failure Management},
  journal = {ACM Transactions on Intelligent Systems and Technology}, volume = {12}, number = {6}, pages = {81:1--81:45}, year = {2021}, doi = {10.1145/3483424}
}

@inproceedings{orchard1972missing,
  author = {Orchard, Terence and Woodbury, Max A.}, title = {A missing information principle: theory and applications},
  booktitle = {Proceedings of the Sixth Berkeley Symposium on Mathematical Statistics and Probability, Volume 1: Theory of Statistics}, pages = {697--716}, publisher = {University of California Press}, year = {1972}
}

@article{owen2020universal,
  author = {Owen, Jeremy A. and Gingrich, Todd R. and Horowitz, Jordan M.}, title = {Universal Thermodynamic Bounds on Nonequilibrium Response with Biochemical Applications},
  journal = {Physical Review X}, volume = {10}, number = {1}, pages = {011066}, year = {2020}, doi = {10.1103/PhysRevX.10.011066}
}

@article{page1954continuous,
  author = {Page, E. S.}, title = {Continuous inspection schemes}, journal = {Biometrika}, volume = {41}, number = {1/2}, pages = {100--115}, year = {1954}, doi = {10.2307/2333009}
}

@book{pavliotis2008multiscale,
  author = {Pavliotis, Grigorios A. and Stuart, Andrew M.}, title = {Multiscale Methods: Averaging and Homogenization}, publisher = {Springer}, address = {New York}, year = {2008},
  series = {Texts in Applied Mathematics}, doi = {10.1007/978-0-387-73829-1}
}

@article{perezhernandez2013identification,
  author = {P{\'e}rez-Hern{\'a}ndez, Guillermo and Paul, Fabian and Giorgino, Toni and De Fabritiis, Gianni and No{\'e}, Frank}, title = {Identification of slow molecular order parameters for {M}arkov model construction},
  journal = {The Journal of Chemical Physics}, volume = {139}, pages = {015102}, year = {2013}, number = {1}, doi = {10.1063/1.4811489}
}

@article{polettini2017effective,
  author = {Polettini, Matteo and Esposito, Massimiliano}, title = {Effective Thermodynamics for a Marginal Observer},
  journal = {Physical Review Letters}, volume = {119}, number = {24}, pages = {240601}, year = {2017}, doi = {10.1103/PhysRevLett.119.240601}
}

@article{polettini2019effective,
  author = {Polettini, Matteo and Esposito, Massimiliano}, title = {Effective Fluctuation and Response Theory},
  journal = {Journal of Statistical Physics}, volume = {176}, number = {1}, pages = {94--168}, year = {2019}, doi = {10.1007/s10955-019-02291-7}
}

@article{politis1994stationary,
  author = {Politis, Dimitris N. and Romano, Joseph P.}, title = {The Stationary Bootstrap}, journal = {Journal of the American Statistical Association}, volume = {89}, number = {428}, pages = {1303--1313}, year = {1994}, doi = {10.1080/01621459.1994.10476870}
}

@article{politis2004automatic,
  author = {Politis, Dimitris N. and White, Halbert}, title = {Automatic Block-Length Selection for the Dependent Bootstrap}, journal = {Econometric Reviews}, volume = {23}, number = {1}, pages = {53--70}, year = {2004}, doi = {10.1081/ETC-120028836}
}

@article{prinz2011markov,
  author = {Prinz, Jan-Hendrik and Wu, Hao and Sarich, Marco and Keller, Bettina and Senne, Martin and Held, Martin and Chodera, John D. and Sch{\"u}tte, Christof and No{\'e}, Frank}, title = {{M}arkov models of molecular kinetics: Generation and validation},
  journal = {The Journal of Chemical Physics}, volume = {134}, pages = {174105}, year = {2011}, number = {17}, doi = {10.1063/1.3565032}
}

@article{prost2009generalized,
  author = {Prost, J. and Joanny, J.-F. and Parrondo, J. M. R.}, title = {Generalized Fluctuation-Dissipation Theorem for Steady-State Systems},
  journal = {Physical Review Letters}, volume = {103}, number = {9}, pages = {090601}, year = {2009}, doi = {10.1103/PhysRevLett.103.090601}
}

@article{puglisi2010entropy,
  author = {Puglisi, Andrea and Pigolotti, Simone and Rondoni, Lamberto and Vulpiani, Angelo}, title = {Entropy production and coarse graining in {M}arkov processes},
  journal = {Journal of Statistical Mechanics: Theory and Experiment}, pages = {P05015}, year = {2010}, eprint = {1002.4520}, archivePrefix = {arXiv}, volume = {2010}, number = {05}, doi = {10.1088/1742-5468/2010/05/P05015}
}

@article{ramsey1996time,
  author = {Ramsey, James B. and Rothman, Philip}, title = {Time Irreversibility and Business Cycle Asymmetry}, journal = {Journal of Money, Credit and Banking}, volume = {28}, number = {1}, pages = {1--21}, year = {1996}
}

@article{rao2016nonequilibrium,
  author = {Rao, Riccardo and Esposito, Massimiliano}, title = {Nonequilibrium Thermodynamics of Chemical Reaction Networks: Wisdom from Stochastic Thermodynamics},
  journal = {Physical Review X}, volume = {6}, pages = {041064}, year = {2016}, number = {4}, doi = {10.1103/PhysRevX.6.041064}
}

@article{rouska2025equilibrium,
  author = {Rouska, Nefeli-Marina and Casamayor Pujol, V{\'\i}ctor and Magrans de Abril, Ildefons and Dustdar, Schahram}, title = {Equilibrium-Driven Antifragility in Computing Continuum Systems},
  journal = {IEEE Internet Computing}, volume = {29}, number = {5}, pages = {55--64}, year = {2025}, doi = {10.1109/MIC.2025.3597479}
}

@article{sedlak2024equilibrium,
  author = {Sedlak, Boris and Casamayor Pujol, V{\'\i}ctor and Donta, Praveen Kumar and Dustdar, Schahram}, title = {Equilibrium in the Computing Continuum through Active Inference},
  journal = {Future Generation Computer Systems}, volume = {160}, pages = {92--108}, year = {2024}, doi = {10.1016/j.future.2024.05.056}
}

@article{seifert2010fluctuation,
  author = {Seifert, Udo and Speck, Thomas}, title = {Fluctuation-dissipation theorem in nonequilibrium steady states},
  journal = {Europhysics Letters}, volume = {89}, number = {1}, pages = {10007}, year = {2010}, doi = {10.1209/0295-5075/89/10007}
}

@article{shiraishi2022time,
  author = {Shiraishi, Naoto}, title = {Time-Symmetric Current and Its Fluctuation Response Relation around Nonequilibrium Stalling},
  journal = {Physical Review Letters}, volume = {129}, pages = {020602}, year = {2022}, doi = {10.1103/PhysRevLett.129.020602}
}

@article{simon1961aggregation,
  author = {Simon, Herbert A. and Ando, Albert}, title = {Aggregation of Variables in Dynamic Systems}, journal = {Econometrica}, volume = {29}, number = {2}, pages = {111--138}, year = {1961}
}

@article{soldani2022anomaly,
  author = {Soldani, Jacopo and Brogi, Antonio}, title = {Anomaly Detection and Failure Root Cause Analysis in (Micro) Service-Based Cloud Applications: A Survey},
  journal = {ACM Computing Surveys}, volume = {55}, number = {3}, pages = {59:1--59:39}, year = {2022}, doi = {10.1145/3501297}
}

@article{spiegelhalter1993bayesian,
  author = {Spiegelhalter, David J. and Dawid, A. Philip and Lauritzen, Steffen L. and Cowell, Robert G.}, title = {{B}ayesian analysis in expert systems},
  journal = {Statistical Science}, volume = {8}, number = {3}, pages = {219--247}, year = {1993}, doi = {10.1214/ss/1177010888}
}

@article{spinner2015evaluating,
  author = {Spinner, Simon and Casale, Giuliano and Brosig, Fabian and Kounev, Samuel}, title = {Evaluating approaches to resource demand estimation},
  journal = {Performance Evaluation}, volume = {92}, pages = {51--71}, year = {2015}, doi = {10.1016/j.peva.2015.07.005}
}

@article{steuber2012testing,
  author = {Steuber, Tara L. and Kiessler, Peter C. and Lund, Robert}, title = {Testing for Reversibility in {M}arkov Chain Data},
  journal = {Probability in the Engineering and Informational Sciences}, volume = {26}, number = {4}, pages = {593--611}, year = {2012}, doi = {10.1017/S0269964812000228}
}

@book{sugiyama2012density,
  author = {Sugiyama, Masashi and Suzuki, Taiji and Kanamori, Takafumi}, title = {Density Ratio Estimation in Machine Learning}, publisher = {Cambridge University Press}, address = {Cambridge}, year = {2012}
}

@inproceedings{tariq2008answering,
  author = {Tariq, Mukarram and Zeitoun, Amgad and Valancius, Vytautas and Feamster, Nick and Ammar, Mostafa}, title = {Answering what-if deployment and configuration questions with {WISE}},
  booktitle = {Proceedings of the ACM SIGCOMM 2008 Conference on Data Communication}, pages = {99--110}, year = {2008}, doi = {10.1145/1402958.1402971}
}

@article{teza2020exact,
  author = {Teza, Gianluca and Stella, Attilio L.}, title = {Exact Coarse Graining Preserves Entropy Production out of Equilibrium},
  journal = {Physical Review Letters}, volume = {125}, pages = {110601}, year = {2020}, number = {11}, doi = {10.1103/PhysRevLett.125.110601}
}

@misc{thommen2026limits,
  author = {Thommen, Quentin}, title = {Limits of Inferring Parametric Response from Single-Condition Trajectories in Stochastic Reaction Networks},
  year = {2026}, eprint = {2609.17031}, howpublished = {arXiv:2609.17031}, archivePrefix = {arXiv}
}

@incollection{thurgood2018alerting,
  author = {Thurgood, Steven and Frame, Jessie and Lenton, Alex and Quinito, Carmela and Tolchanov, Anton and Trdin, Nejc}, title = {Alerting on {SLO}s},
  booktitle = {The Site Reliability Workbook: Practical Ways to Implement SRE}, editor = {Beyer, Betsy and Murphy, Niall Richard and Rensin, David K. and Kawahara, Kent and Thorne, Stephen}, publisher = {O'Reilly Media}, year = {2018}
}

@article{tsoucas1987interchangeability,
  author = {Tsoucas, Pantelis and Walrand, Jean}, title = {On the interchangeability and stochastic ordering of $\cdot$/{M}/1 queues in tandem},
  journal = {Advances in Applied Probability}, volume = {19}, number = {2}, pages = {515--520}, year = {1987}, doi = {10.2307/1427433}
}

@article{wald1945sequential,
  author = {Wald, Abraham}, title = {Sequential Tests of Statistical Hypotheses}, journal = {The Annals of Mathematical Statistics}, volume = {16}, number = {2}, pages = {117--186}, year = {1945}, doi = {10.1214/aoms/1177731118}
}

@incollection{warner2018canarying,
  author = {Warner, Alec and Davidovi{\v{c}}, {\v{S}}tefan and Hidalgo, Alex and Beyer, Betsy and Smith, Kyle and Duftler, Matt}, title = {Canarying Releases},
  booktitle = {The Site Reliability Workbook: Practical Ways to Implement SRE}, editor = {Beyer, Betsy and Murphy, Niall Richard and Rensin, David K. and Kawahara, Kent and Thorne, Stephen}, publisher = {O'Reilly Media}, year = {2018}
}

@article{weber1979interchangeability,
  author = {Weber, Richard R.}, title = {The interchangeability of $\cdot$/{M}/1 queues in series},
  journal = {Journal of Applied Probability}, volume = {16}, number = {3}, pages = {690--695}, year = {1979}, doi = {10.2307/3213100}
}

@article{weiss1975time,
  author = {Weiss, Gideon}, title = {Time-reversibility of linear stochastic processes}, journal = {Journal of Applied Probability}, volume = {12}, number = {4}, pages = {831--836}, year = {1975}, doi = {10.2307/3212735}
}

@article{wilson1927probable,
  author = {Wilson, Edwin B.}, title = {Probable Inference, the Law of Succession, and Statistical Inference}, journal = {Journal of the American Statistical Association}, volume = {22}, number = {158}, pages = {209--212}, year = {1927}, doi = {10.1080/01621459.1927.10502953}
}

@article{woodside1995stochastic,
  author = {Woodside, C. Murray and Neilson, John E. and Petriu, Dorina C. and Majumdar, Shikharesh}, title = {The Stochastic Rendezvous Network Model for Performance of Synchronous Client-Server-like Distributed Software},
  journal = {IEEE Transactions on Computers}, volume = {44}, number = {1}, pages = {20--34}, year = {1995}, doi = {10.1109/12.368012}
}

@article{wu2025parameter,
  author = {Wu, Bingjie and Jia, Chen}, title = {Parameter Inference and Nonequilibrium Identification for {M}arkov Networks Based on Coarse-Grained Observations},
  journal = {Physical Review Letters}, volume = {134}, pages = {087103}, year = {2025}, eprint = {2406.19586}, archivePrefix = {arXiv}, number = {8}, doi = {10.1103/PhysRevLett.134.087103}
}

@book{yin2013continuous,
  author = {Yin, G. George and Zhang, Qing}, title = {Continuous-Time {M}arkov Chains and Applications: A Two-Time-Scale Approach}, edition = {2nd}, publisher = {Springer}, address = {New York}, year = {2013},
  series = {Stochastic Modelling and Applied Probability}, volume = {37}, doi = {10.1007/978-1-4614-4346-9}
}

@article{zheng2025unified,
  author = {Zheng, Jiming and Lu, Zhiyue}, title = {Unified linear fluctuation-response theory arbitrarily far from equilibrium},
  journal = {Physical Review E}, volume = {112}, pages = {064103}, year = {2025}, doi = {10.1103/rgys-zxgf}, eprint = {2501.01050}, archivePrefix = {arXiv}
}

@misc{zheng2026memory,
  author = {Zheng, Jiming and Lu, Zhiyue}, title = {The Memory Hidden in Response Fluctuations: Trajectory-Level Fluctuation-Response Theory and Inequalities for Non-{M}arkovian Jump Dynamics},
  year = {2026}, eprint = {2608.20328}, howpublished = {arXiv:2608.20328}, archivePrefix = {arXiv}
}

\appendix
\section{Proof of \texorpdfstring{\cref{prop:fdt}}{Proposition 1}}\label{app:fdt}

\paragraph{Regularity.} Let $\bar q$ bound the exit rates $q(x)=\sum_{y\ne x}q(x,y)$ uniformly in $x$ (and in $h$ near $0$). Then $L$ is a bounded operator on $L^2(\pi)$: writing $(Qf)(x)=\sum_{y\ne x}q(x,y)f(y)$, the Cauchy--Schwarz inequality gives $|Qf(x)|^2\le q(x)\sum_{y\ne x}q(x,y)f(y)^2$, and stationarity, $\sum_x\pi(x)q(x,y)=\pi(y)q(y)$, gives $\|Qf\|^2_{L^2(\pi)}\le\bar q^{\,2}\|f\|^2_{L^2(\pi)}$, so $\|L\|\le2\bar q$. The same holds for $L^\dagger$ and, on $\ell^1$, for $L^*$. The semigroups $e^{tL}$ are then norm-continuous and the Duhamel formula below holds; together with the differentiability of $\pi_{\theta+h}$ in $h$ (assumed in \cref{sec:state}) and $A,B\in L^2(\pi)$, this makes every manipulation below legitimate. The mean of $B$ is zero because $\sum_x\pi_{\theta+h}(x)=1$ for all $h$ and the derivative is taken term by term (as it can be for the truncated chains computed here and for the product-form laws of \cref{sec:productform}, where $\partial_h\pi$ is explicit).

\paragraph{Proof.} Write the stationary law of $L_{\theta+h}$ as $\pi+h\,\delta\pi+o(h)$. Stationarity, $(L+hL_1)^*(\pi+h\,\delta\pi)=0$, gives at first order $L^*\delta\pi=-L_1^*\pi$. Since $\delta\pi=\pi B$ with $B=\partial_h\ln\pi$ and $\avg{B}=0$, this reads $L_1^*\pi=-L^*(\pi B)$, i.e.\ the source term of the perturbation is
\[
V:=\frac{L_1^*\pi}{\pi}=-L^\dagger B.
\]
After a step of size $h$ at time $0$, the deviation of the law at time $\tau$ is $h\int_0^\tau e^{(\tau-s)L^*}L_1^*\pi\,\dd s+o(h)$ (Duhamel), so the response of $\avg{A}$ is
\[
R_A(\tau)=\int_0^\tau\avg{A(X_u)\,V(X_0)}\,\dd u.
\]
On the other hand, $\frac{\dd}{\dd s}C_{A,B}(s)=\avg{A(X_s)\,(L^\dagger B)(X_0)}=-\avg{A(X_s)\,V(X_0)}$. Integrating,
\[
R_A(\tau)=-\int_0^\tau\frac{\dd}{\dd s}C_{A,B}(s)\,\dd s=C_{A,B}(0)-C_{A,B}(\tau).\qquad\qed
\]

\section{Numerical setup and reproduction}\label{app:numerics}

\paragraph{Models.} Two-stage tandem with $\lambda=1$, $\mu_1=2$, $\mu_2=1.6$. Model J: unlimited buffers, truncated at 70 jobs per stage. Model B: a buffer of 3 at stage 2 with blocking after service; stage 1 truncated at 70 ($P(n_1=70)=9\times10^{-12}$). Model B2 (\cref{sec:autoscaler,sec:finite}): buffer of 2. Multi-stage tandems (\cref{sec:tandem}): truncated at a total of 11 to 24 jobs, with arrivals blocked at the cap; all reported values are stable under the truncation. Model A (\cref{sec:autoscaler}): $\lambda=1.2$, $\mu=1$ per replica, thresholds 8 and 1, truncated at $n=160$ (and $n=240$ for the lumped-detailed-balance check). Stationary laws by a sparse linear solve; time-dependent quantities by sparse matrix-exponential actions (scipy \texttt{expm\_multiply}); derivatives with respect to controls by central differences with $h=10^{-5}$ ($10^{-6}$ for the score of model A in \texttt{autoscaler\_lumped\_db.py}). Exact step responses come from the perturbed chain started from the unperturbed stationary law. Simulations use exact uniformisation from a stationary initial state.

\paragraph{Scripts.} Python~3 with numpy and scipy (matplotlib for the figure). The exact checks run in seconds to a few minutes on a laptop; the finite-sample study takes about one to two hours of single-core time for the four models. Stored outputs, including the per-replicate statistics of \cref{sec:finite}, are part of the repository. \Cref{tab:scripts} lists the scripts.

\begin{table}[ht]
\centering
\caption{Scripts of the repository and the parts of the paper they produce.}
\label{tab:scripts}
\footnotesize
\begin{tabularx}{\textwidth}{@{}>{\raggedright\arraybackslash}p{6.1cm}X>{\raggedright\arraybackslash}p{2.3cm}@{}}
\toprule
Script & Content & Where\\
\midrule
\texttt{exact/verify\_tloe.py} & Product form, tests of \cond{F} and \cond{R}, FDT predictions with $B=N,n_1,n_2,-n_1$ and the exact conjugate, reciprocity & \cref{sec:chains}, \cref{tab:conjugate,tab:blocking}\\
\texttt{exact/verify\_independence\_and\_kl.py} & Three-time symmetry with positive control (model J), stage-permutation invariance, \cond{R} without \cond{F}, small-$\delta$ KL law, truncation check & \cref{sec:independence,sec:blocking,sec:kl}\\
\texttt{exact/kl\_bridge.py} & KL between observed and tilted laws of $N$, models J and B & \cref{sec:kl}\\
\texttt{exact/identifiability.py} & FDT with the top-level conjugate $B^Y$, models J and B & \cref{sec:testF}\\
\texttt{exact/eigen\_check.py} & Generalised eigenvalues of the lagged moments, models J and B & \cref{sec:blocking}\\
\texttt{exact/extra\_checks.py} & Capacity rows of \cref{tab:conjugate}; three-time asymmetry for model B; Chapman--Kolmogorov test and implied time scales of $N$; decay time scale of $C_{NN}$; KL extrapolation; error conventions of \cref{sec:fastcontrols} & \cref{sec:state,sec:consequences,sec:chains,sec:kl,sec:fastcontrols}\\
\texttt{exact/weber\_conditional\_check.py}, \texttt{exact/positive\_control\_erlang.py} & Fixed-arrival interchangeability (exact) and its Erlang-2 positive control & \cref{sec:tandem}\\
\texttt{exact/kstage.py}, \texttt{exact/stage\_controls.py} & $K$-stage tandems; block interchangeability (\cref{lem:block}); reversibility of $(N,n_j)$ & \cref{sec:tandem}, \cref{tab:stage}\\
\texttt{exact/autoscaler.py}, \texttt{exact/autoscaler\_table.py}, \texttt{exact/autoscaler\_sufficiency.py}, \texttt{exact/compare\_metrics.py} & Model A: \cond{F}, \cond{R}, prediction errors across $\alpha$; minimal sufficient statistic; the same metrics for models J, B, B2 & \cref{sec:autoscaler}, \cref{tab:autoscaler}\\
\texttt{exact/autoscaler\_lumped\_db.py} & Model A: lumped detailed balance \eqref{eq:lumpedDB} on the level sets of the score, $\alpha\in\{0.003,0.03,0.1,1,100\}$, truncation 160 and 240 & \cref{sec:autoscaler}\\
\texttt{exact/cross\_scale\_chain\_rule.py} & Cross-scale rule on a modulated M/M/1 queue & \cref{sec:fastcontrols}\\
\texttt{exact/ps\_vs\_fcfs\_erlang.py} & Single queue with Erlang-2 service, FCFS against processor sharing: sufficiency of $n$ for the load and accuracy of the prediction & \cref{sec:limitations}\\
\texttt{exact/blocking\_minimal\_statistic.py} & Model B: minimal sufficient statistic for the load; bias of the prediction with $B=N$ against the buffer size & \cref{sec:blocking}\\
\texttt{exact/fig\_evidence.py} & \Cref{fig:evidence} & \cref{sec:chains}\\
\texttt{finite\_sample/exact.py}, \texttt{finite\_sample/sim.py}, \texttt{finite\_sample/models.py}, \texttt{finite\_sample/autoscaler.py} & Exact references and exact simulators for J, B3, B2, A & \cref{sec:finite}\\
\texttt{finite\_sample/study.py} & Prediction error, T-R and baseline tests, sensitivity runs (1{,}000 replicates per model) & \cref{sec:finite}\\
\texttt{finite\_sample/summarise.py} & \Cref{tab:pred,tab:tr,tab:sens} with confidence intervals & \cref{sec:finite-pred,sec:finite-tr,sec:sensitivity}\\
\texttt{finite\_sample/tf\_power.py}, \texttt{finite\_sample/tf\_power\_A.py} & Cost of a dynamic T-F test with the hypothesised and the estimated conjugates (models B3, B2; model A) & \cref{sec:tfcost}, \cref{tab:tfcost}\\
\texttt{finite\_sample/tr\_signal\_strength.py} & Exact size of the time-asymmetry signal & \cref{sec:finite}\\
\bottomrule
\end{tabularx}
\end{table}

\end{document}